\documentstyle[eqsecnum,preprint,aps]{revtex}
                                                             
\begin{document}      

\draft
\preprint{TPR-96-15}

\title{Equal-Time Hierarchies for Quantum Transport Theory}

\author{Pengfei Zhuang\cite{home}}

\address{Gesellschaft f\"ur Schwerionenforschung, Theory Group, \\
         P.O.Box 110552, D-64220 Darmstadt, Germany}

\author{Ulrich Heinz} 

\address{Institut f\"ur Theoretische Physik, Universit\"at Regensburg,\\ 
         D-93040 Regensburg, Germany } 

\date{\today}

\maketitle 

\begin{abstract}

We investigate in the equal-time formalism the derivation and truncation
of infinite hierarchies of equations of motion for the energy moments of 
the covariant Wigner function. From these hierarchies we then extract
kinetic equations for the physical distribution functions which are 
related to low-order energy moments, and show how to determine 
the higher order moments in terms of these lowest order ones. We apply 
the general formalism to scalar and spinor QED with classical background 
fields and compare with the results derived from the three-dimensional 
Wigner transformation method. 

\end{abstract} 


\section{Introduction}
\label{sec1}

Transport theory \cite{EH} based on the Wigner operator is extensively 
used to describe the formation and evolution of highly excited nuclear 
matter produced in relativistic heavy ion collisions. The Wigner 
operator can be defined in 4-dimensional \cite{CZ,He,EGV1,EGV2,VGE} or 
3-dimensional \cite{BGR,BGG} momentum space, which we denote by $\hat 
{\cal W}(x,p)$ and $\hat W(x,\bbox{p})$, respectively. Correspondingly, 
there are two formulations for the phase-space structure of any field. 
Either of these two formulations has its advantages and disadvantages. 
Besides its manifest Lorentz covariance (which is very useful from a 
technical point of view), another characteristic feature of 
the 4-dimensional formulation for QCD \cite{CZ,He,EGV1,EGV2} and QED 
\cite{VGE} is that the quadratic kinetic equation can be split up 
naturally into a transport and a constraint equation. The 
complementarity of these two ingredients is essential for a physical 
understanding of quantum kinetic theory \cite{Hen}. In the classical 
limit, these two equations reduce to the Vlasov and mass-shell 
equations, respectively. The main advantage of the 3-dimensional 
approach \cite{BGR,BGG} is that it is easier to set up as an initial 
value problem: one can directly compute the initial value of the 
Wigner operator from the corresponding field operators at the same 
time. In the covariant frame this is not possible since the covariant 
Wigner operator is defined as a 4-dimensional Wigner transform of the 
density matrix and thus includes an integration over time. Hence in 
this approach the initial condition for the Wigner operator at very 
early times must be constructed phenomenologically. Some true quantum 
problems like pair production \cite{Sch} in a strong external field 
have thus so far been solved only in the 3-dimensional (or equal-time) 
formulation \cite{BGR,BE}.  

One way \cite{BGR} to obtain equal-time kinetic equations which 
parallels the procedure in the covariant formulation is to Wigner 
transform the equation of motion for the equal-time density operator 
$\hat \Phi(\bbox{x},\bbox{y},t)$. For spinor QED this procedure results 
in the BGR equations \cite{BGR} for the equal-time Wigner functions. In 
Ref.~\cite{ZH1} we suggested a different derivation which is based on 
taking the energy average of the covariant kinetic equations in the 
4-dimensional formulation. It exploits the fact that the equal-time 
Wigner function is the energy average (i.e. zeroth order energy 
moment) of the covariant one. With this method we showed for spinor 
QED that the direct energy average of the covariant kinetic equations 
leads, in addition to the BGR transport equations for the spinor 
components of the equal-time Wigner function, also to a second group 
of constraint equations which couple the equal-time Wigner function to 
the first order energy moment of the covariant one. In the classical 
($\hbar\to 0$) limit, these additional equations provide essential 
constraints on the equal-time Wigner function and allow to reduce the 
number of independent distribution functions by a factor of two 
\cite{ZH1}. In the general quantum case, the additional equations 
determine the time evolution of the energy distribution function which 
in general can not be expressed in terms of the equal-time Wigner 
function. In this sense the BGR equations do not provide a complete 
set of equal-time kinetic equations.  

As we will discuss in this paper, this incompleteness has a more 
general aspect. As just mentioned, the equal-time Wigner operator is 
related to the covariant one by \cite{ZH1} 
 \begin{equation}
 \label{W3}
   \hat W(x,\bbox{p}) = \int dE\, \hat {\cal W}(x,\bbox{p},E)\, ,
 \end{equation}
where we wrote $p=(E,\bbox{p})$, $E$ independent of $\bbox{p}$. As such 
it is only the lowest member of an infinite hierarchy of energy 
moments of the covariant Wigner operator: 
 \begin{equation}
 \label{moment1}
   \hat W_j(x,\bbox{p}) = \int dE\, E^j\, \hat {\cal W}(x,\bbox{p},E)\, ,
   \qquad   j=0,1,2,\cdots\, , 
 \end{equation}
with $\hat W_0(x,\bbox{p}) \equiv \hat W(x,\bbox{p})$. Therefore, to 
set up a complete equal-time transport theory which contains the same 
amount of information as the covariant theory one needs dynamical 
equations for all the energy moments. Any covariant kinetic equation 
will thus correspond to an infinite hierarchy of coupled kinetic 
equations for its energy moments, i.e. for the equal-time Wigner 
operators $\hat W_j(x,\bbox{p})$.  

This infinite hierarchy only exists for genuine quantum problems where 
the energy can exhibit quantum fluctuations. In the classical limit, 
the covariant Wigner operator satisfies the mass-shell constraint 
$p^2=E^2-\bbox{p}^2=m^2$, and the energy dependence of the covariant 
Wigner function thus degenerates to two delta-functions at $E = \pm 
E_p = \pm \sqrt{m^2+\bbox{p}^2}$. The equal-time Wigner operator $\hat 
W(x,\bbox{p})$ then splits into a positive and a negative frequency 
component, 
 \begin{equation}
 \label{class1}
   \hat W(x,\bbox{p}) = 
   \hat W^+(x,\bbox{p}) + \hat W^-(x,\bbox{p})\, ,
   \qquad (\hbar \to 0)
 \end{equation}
and all energy moments can be expressed algebraically in terms of 
these as 
 \begin{eqnarray}
 \label{class2}
   \hat W_j(x,\bbox{p}) &=& \hat W_j^+(x,\bbox{p})+\hat W_j^-(x,\bbox{p})
                           \nonumber\\
                       &=& E_p^j\, \hat W^+(x,\bbox{p}) 
                           + (-E_p)^j \, \hat W^-(x,\bbox{p})\, , 
                           \quad j=0,1,2,\dots
                           \quad (\hbar \to 0).
 \end{eqnarray}
The solution of the equal-time kinetic equations for $\hat W(x,\bbox{p})$
thus also determines the dynamics of all higher energy moments. Thus, 
in the classical limit, a simple zeroth order energy average of the 
covariant kinetic theory yields a complete equal-time kinetic theory.

In the general quantum case, the higher order energy moments $\hat 
W_j(x,\bbox{p})$, $j\geq 1$, contain genuine additional information 
and can no longer be expressed algebraically through the equal-time 
Wigner operator $\hat W(x,\bbox{p})$. This means that in principle in 
the equal-time formulation we are stuck with the problem of solving 
an infinite hierarchy of coupled equations. Actually, there are two 
such hierarchies, one resulting from the covariant transport equation
(``transport hierarchy''), the other arising from the generalized 
mass-shell constraint (``constraint hierarchy''). In practice this
raises the problem of truncating the hierarchy in a physically sensible 
way. Since only the low-order energy moments of the covariant Wigner 
function have an intuitive physical interpretation, it turns out that 
physics itself suggests an appropriate truncation scheme. We will show 
that the hierarchies of moment equations are structured in such a way that
the first few low-order moments form a finite and closed subgroup of
equations which can be solved as an initial value problem, and that 
(surprisingly) all the higher order moments can be derived from these
low-order moments recursively using only the constraint hierarchy, i.e.
without solving any additional equations of motion. The equations
from the transport hierarchy for the higher order moments are redundant.

We first discuss on a general basis, starting from the covariant 
approach, the derivation and truncation of equal-time hierarchies 
of kinetic equations. For illustration we then consider in full
generality the case of a transport theory for scalar fields with 
arbitrary scalar potentials. For this case everything can be worked 
out explicitly to arbitrary order of the moments. We give the subgroup 
of equations which fully characterize the first few low-order moments, 
prove the independence and redundance of the transport equations for 
the higher order moments outside this subgroup, and obtain from the 
constraint hierarchy explicit expressions for all the higher order 
moments in terms of the solutions of the low-order subgroup. We then 
apply the general formalism to scalar and spinor QED. Here the equations
have a more complicated structure, and we restrict our attention to the
closed subgroup of equations for the lowest order moments, discussing 
the redundance of the transport equations for the higher order moments 
and their recursive determination through the constraint hierarchy only 
for the first moments outside the closed subgroup of equations for the 
low-order moments. We will compare our results with the kinetic equations 
for the equal-time distribution functions obtained previously in 
Refs.~\cite{BGR,BGG,ZH1}. Our final result will be a complete set of 
kinetic equations which can be implemented numerically as an initial 
value problem.

\section{General Formalism}
\label{sec2}

The 4-dimensional Wigner transform of the equation of motion for the 
covariant density operator leads to a complex, Lorentz covariant 
kinetic equation for the Wigner operator. It couples the one-body 
Wigner operator to two-body correlations \cite{EH}, which in turn 
satisfy an equation which couples them to three-body terms, and so on. 
After taking an ensemble average this generates the so-called BBGKY 
hierarchy \cite{GLW} for the $n$-body Wigner functions. A popular way 
to get a closed kinetic equation for the one-body Wigner function 
(i.e. the ensemble average of the one-body Wigner operator) is to truncate
the BBGKY hierarchy at the one-body level, by factorizing the two-body
Wigner functions in the Hartree approximation. So far most applications
of quantum transport theory have employed this approximation, and in 
the following we will also restrict ourselves to it. For us the mean field
approximation provides a crucial simplification, and at present it is 
not obvious to us how to generalize our results in order to include 
correlations and collision terms.

For a scalar field in mean field approximation the complex equation for 
the self-adjoint scalar Wigner function can be separated into two 
independent real equations \cite{ZH1}:
 \begin{mathletters}
 \label{scalar}
 \begin{eqnarray}
 \label{scalar1}
   \hat G(x,p) \, {\cal W}(x,p) &=& 0\, ;
 \\
 \label{scalar2}
   \hat F(x,p) \, {\cal W}(x,p) &=& 0\, .
 \end{eqnarray}
 \end{mathletters}
The first equation corresponds to a generalized Vlasov equation; after 
performing the energy average it generates a hierarchy of transport 
equations for the energy moments $W_j(x,\bbox{p})$ (``transport 
hierarchy'').  The second equation is a generalized mass-shell 
constraint; it generates a hierarchy of non-dynamic constraint 
equations (``constraint hierarchy''). Equations with the structure 
given in (\ref{scalar}) will be the starting point for our discussion 
of scalar field theories in Secs.~\ref{sec2c} and \ref{sec3a}. Factors 
of $p_\mu$ in the dynamical operators $\hat G(x,p)$ and $\hat F(x,p)$ 
arise from the Wigner transformation of the partial derivative 
$\partial_\mu$ in the Klein-Gordon equation. Since the latter contains 
at most second order time derivatives, at most two powers of $p_0$ 
occur. In fact, $\hat G(x,p)$ is linear in $p_0$ while $\hat F(x,p)$ 
is quadratic in $p_0$. This will be important below (see 
Sec.~\ref{sec2b}).  

For spinor fields the covariant Wigner function is a $4\times 4$ 
matrix in spinor space which for a physical interpretation must be 
decomposed into its 16 spinor components ${\cal W}^s$. In this way the 
complex kinetic equation for the Wigner function matrix is split into 
$32$ independent real equations for the self-adjoint spinor components 
\cite{VGE,ZH1}. These equations can be further divided into two 
subgroups according to their anticipated structure after performing 
the energy average:
 \begin{mathletters}
 \label{spinor}
 \begin{eqnarray}
 \label{spinor1}
   \sum_{s'=1}^{16} \hat G^{ss'}(x,p) \, {\cal W}^{s'}(x,p) &=& 0\, ,
 \\
 \label{spinor2}
   \sum_{s'=1}^{16} \hat F^{ss'}(x,p) \, {\cal W}^{s'}(x,p) &=& 0\, ,
   \quad (s=1,2,\dots,16).
 \end{eqnarray}
 \end{mathletters}
The first subgroup leads to equations containing only first order time 
derivatives and thus generates 16 hierarchies of transport equations 
for the energy moments of the spinor components (``transport 
hierarchies''). The other subgroup which involves both first and 
second order time derivatives leads to a set of 16 hierarchies of 
constraint equations (``constraint hierarchies'') for the equal-time 
moments of the spinor components. For the lowest energy moments, the 
spinor components $W^s(x,\bbox{p})$ of the equal-time Wigner function, 
the details of this procedure were worked out in 
Ref.~\cite{ZH1}, and we will use these results in Sec.~\ref{sec3c}.  
Since the original Dirac equation is linear in the time derivative, 
the dynamical operators $\hat G^{ss'}(x,p)$ and $\hat F^{ss'}(x,p)$ 
contain at most single powers of $p_0$. In fact, the operators $\hat 
G^{ss'}(x,p)$ are independent of $p_0$.  

\subsection{Hierarchy of energy moments}
\label{sec2a}

In this subsection we will concentrate for simplicity on a single 
covariant kinetic equation of the generic form
 \begin{equation}
 \label{generic}
   \hat G(x,p) \, {\cal W}(x,p) = 0\, ,
 \end{equation}
where $\hat G(x,p)$ contains at most two powers of $p_0$ and of the 
space-time derivative operator $\partial_\mu$, but an arbitrary number 
of derivatives with respect to the momentum space coordinates (see 
Appendix~\ref{appc}). We will return to the full set of equations 
(\ref{scalar}) resp. (\ref{spinor}) in the following Sections. 

We begin by decomposing the energy dependence of the Wigner function 
into a basis of orthogonal polynomials $h_j(E)$: 
 \begin{equation}
 \label{expan}
   {\cal W}(x,\bbox{p},E) =  
   \sum_{j=0}^\infty w_j(x,\bbox{p})\, h_j(E)\ .
 \end{equation}
The expansion coefficients $w_j(x,\bbox{p})$ are defined in the 
equal-time phase space. Using the orthonormality relation 
 \begin{equation}
 \label{ortho}
   \int d\mu(E)\, h_i(E)\, h_j(E) = \delta_{ij}\ ,
 \end{equation}
where $d\mu(E)$ is the appropriate integration measure associated with 
the chosen set of polynomials $h_j$, the equal-time components 
$w_j(x,\bbox{p})$ can be related to energy moments of the covariant 
Wigner function constructed with the basis functions $h_j(E)$: 
 \begin{equation}
 \label{moment2}
   w_j(x,\bbox{p}) = \int d\mu(E)\, h_j(E)\, {\cal W}(x,\bbox{p},E)\ .
 \end{equation}
If the system has finite total energy the covariant spinor components 
must vanish in the limit $E \to \pm \infty$. We will assume that they
vanish at infinite energy faster than any power of $E$ such that for 
any combination of integers $i,j,m,n\geq 0$, we have
 \begin{equation}
 \label{surface}
     \int d\mu(E) {\partial\over \partial E}\left( 
       {\partial^n\over\partial E^n}\left(h_i(E)E^m\right)
       {\partial^j\over\partial E^j} {\cal W}(x,\bbox{p},E)\right) = 0\ .
 \end{equation}
With exponential accuracy we may therefore restrict the energy 
integration to a finite interval $-\Lambda \leq E \leq \Lambda$.  
Introducing the scaled energy $\omega = E/\Lambda$ we can thus use as 
our set of basis functions the Legendre polynomials
 \begin{equation}
 \label{Legendre}
   h_n(\omega) = \sqrt{{2n+1\over 2}} P_n(\omega)
 \end{equation}
with the trivial measure $d\mu(\omega)=d\omega$ on the interval 
$[-1,1]$. 

As discussed above, the dynamical operator $\hat G(x,p)$ in 
(\ref{generic}) in general contains powers of $E$ up to second order 
and an infinite number of energy derivatives $\partial/\partial E$. In 
terms of the new dimensionless energy variable we may thus write 
 \begin{equation}
 \label{gradient}
   \hat G(x,\bbox{p}, \omega) = 
   \sum_{m=0}^M \sum_{n=0}^\infty \hat G_{mn}(x,\bbox{p})\, \omega^m
   \left({\partial\over \partial \omega}\right)^n\, ,
 \end{equation}
with $M\leq 2$. Substituting this double expansion into equation 
(\ref{generic}), multiplying by $h_i(\omega)$ from the left and 
integrating over energy $\omega$ we obtain 
 \begin{equation}
 \label{kinetic3}
   \sum_{m,n}\hat G_{mn}(x,\bbox{p}) 
   \int^1_{-1} d\omega\, h_i(\omega)\, \omega^m\,
   {\partial^n\over \partial \omega^n} {\cal W}(x,\bbox{p},\omega) = 0\, . 
 \end{equation}
Inserting the expansion (\ref{expan}) of the covariant Wigner function  
${\cal W}(x,\bbox{p},\omega)$, this can be written as
 \begin{equation}
 \label{hierarchy1}
   \sum_{j=0}^\infty \hat H_{ij}(x,\bbox{p})\,  
   w_j(x,\bbox{p}) = 0 \, ,\qquad i=0,1,2,\dots\, ,
 \end{equation}
where
 \begin{equation}
 \label{bigf}
    \hat H_{ij}(x,\bbox{p}) 
    = \sum_{m=0}^M \sum_{n=0}^\infty C_{ij}^{mn}\, \hat G_{mn}(x,\bbox{p})
 \end{equation}
with 
 \begin{equation}
 \label{bigc}
    C_{ij}^{mn} = \int_{-1}^1 d\omega\, 
    h_i(\omega)\, \omega^m\, \partial_\omega^n \, h_j(\omega) \, .
 \end{equation}
It is easy to see from (\ref{bigc}) that
 \begin{equation}
 \label{bigc1}
   C_{ij}^{mn} = 0 \quad \text{for} \quad n>j \quad
   \text{and} \quad i>j+m-n\, .
 \end{equation}
For $j \geq i-m+n$ the coefficients are in general nonzero. Therefore, 
the sum over $j$ in Eq.~(\ref{hierarchy1}) extends over the range 
$j\geq \max[0,i-M]$. For each value of $i$, Eq.~(\ref{hierarchy1}) 
thus contains an infinite number of terms. In this form 
Eqs.~(\ref{hierarchy1}) are thus not practically useful. However, one 
can use the surface condition (\ref{surface}) to rewrite 
Eqs.~(\ref{hierarchy1}) in such a way that each equation contains only 
a finite number of terms. Returning to Eq.~(\ref{kinetic3}) and 
integrating by parts, we can replace the integrand by 
 \begin{eqnarray}
 \label{byparts}
   && h_i(\omega) \, \omega^m \partial^n_\omega {\cal W}(x,\bbox{p},\omega) = 
 \\
   && \sum_{l=0}^{n-1}(-)^l \partial_\omega 
      \left( \partial^l_\omega \Bigl( h_i(\omega)\omega^m \Bigr)
             \partial^{n-l-1}_\omega {\cal W}(x,\bbox{p},\omega) \right)
     + (-)^n \partial^n_\omega \Bigl( h_i(\omega)\omega^m \Bigr)
       {\cal W}(x,\bbox{p},\omega)\, .
 \nonumber
 \end{eqnarray}
The contribution to the integral in (\ref{kinetic3}) from each term 
in the sum is fully cancelled by the surface condition 
(\ref{surface}), and only the last term in (\ref{byparts}) survives.
Inserting it into (\ref{kinetic3}) and using again the expansion 
(\ref{expan}) we find instead of Eqs.~(\ref{hierarchy1}-\ref{bigc}) 
the following set of equations:
 \begin{equation}
 \label{hierarchy2}
      \sum_{j=0}^{i+M}\hat g_{ij}(x,\bbox{p}) \, 
      w_j(x,\bbox{p}) = 0\, , \qquad i=0,1,2,\dots,
 \end{equation}
with
 \begin{equation}
 \label{hierarchy2a}
    \hat g_{ij}(x,\bbox{p}) = \sum_{m=0}^M \sum_{n=0}^{i+m}
      c_{ij}^{mn}\hat G_{mn}(x,\bbox{p})
 \end{equation}
and the coefficients
 \begin{equation}
 \label{hierarchy2b}
    c_{ij}^{mn} = {1\over 2} \sqrt{(2i+1)(2j+1)}
    \int_{-1}^1 d\omega \,
    P_j(\omega) (-\partial_\omega)^n 
    \Bigl( P_i(\omega) \omega^m \Bigr) \, .
 \end{equation}
The latter can be determined recursively from $c^{00}_{ij}=\delta_{ij}$ 
(which results from the orthogonality relation (\ref{ortho})) by using 
the recursion relations for the Legendre polynomials, see 
Appendix~\ref{appa}. Since the nonvanishing coefficients $c_{ij}^{mn}$ 
are now restricted to the domain 
 \begin{equation}
 \label{smallc}
   n \leq i + m \quad \text{and} \quad j \leq i + m - n \, ,
 \end{equation}
the sum over $j$ in (\ref{hierarchy2}) runs now only over the finite 
range $0 \leq j \leq i+M$. The first inequality in (\ref{smallc}) was 
already used in (\ref{hierarchy2a}) to limit the sum over $n$.
  
The $P_i(\omega)$ are polynomials in $\omega$ of order $i$, and thus 
the equal-time components $w_j(x,\bbox{p})$ occurring in 
Eqs.~(\ref{hierarchy2}) are linear combinations of the energy moments 
$W_k(x,\bbox{p})$ of order $k \leq i$ (see Eq.~(\ref{moment1})).  
For each value of $i$, Eq.~(\ref{hierarchy2}) thus provides a relation 
among the first $i+M+1$ energy moments of the covariant Wigner 
function $\hat W(x,\bbox{p},\omega)$ (including the zeroth order moment 
$W(x,\bbox{p}) = \sqrt{2} w_0(x,\bbox{p})$).  As $i$ is 
allowed to run over all positive integers, Eqs.~(\ref{hierarchy2}) 
form an infinite hierarchy of relations among the energy moments of 
the covariant Wigner function. Each covariant equation of the type 
(\ref{generic}) generates its own such hierarchy. Only the full set 
of these infinite hierarchies of moment equations constitutes a 
complete equal-time kinetic description of the system under study.  

\subsection{Truncating the hierarchy}
\label{sec2b}

In order to discuss possible truncation schemes we must return to the 
complete set of covariant kinetic equations. Let us concentrate here 
on the scalar case, Eqs.~(\ref{scalar}), and write down the two 
resulting hierarchies of moment equations as
 \begin{mathletters}
 \label{hier}
 \begin{eqnarray}
 \label{hier1}
   \sum_{j=0}^{i+M} \hat g_{ij}(x,\bbox{p}) w_j(x,\bbox{p}) &=& 0\, ,\\
 \label{hier2}
   \sum_{j=0}^{i+M+1} \hat f_{ij}(x,\bbox{p}) w_j(x,\bbox{p}) &=& 0\, ,
 \quad (i=0,1,2,\dots).
 \end{eqnarray}
 \end{mathletters}
In writing down the upper limits of the sums we already used that $\hat 
F(x,p)$ in (\ref{scalar1}) contains one power of $p_0$ more than $\hat 
G(x,p)$ in (\ref{scalar1}). For the scalar field case one has $M=1$. 
For the spinor case one obtains from (\ref{spinor}) a similar set of 
equations with $M=0$.  
 
Let us now try to truncate these hierarchies for the moments $w_j$ at 
some order $j_{\rm max}$. The equations from the ``transport
hierarchy'' (\ref{hier1}) with hierarchy index $i\leq I_t$ involve 
all moments $w_j$ with $0\leq j\leq I_t+M$, i.e. the first $I_t+M+1$ 
moments (including the lowest moment with index 0). Similarly, the 
equations from the ``constraint hierarchy'' (\ref{hier2}) with 
hierarchy index $i\leq I_c$ involve the first $I_c+M+2$ moments 
$0\leq j \leq I_c+M+1$. For a closed set of equations both hierarchies 
must be truncated at the same order $j_{\rm max}$, i.e. we must have
 \begin{equation}
 \label{condition1}
   I_t+M = I_c+M+1 = j_{\rm max}\, .
 \end{equation}
Truncating in this way we are left with $I_t+1$ equations from the 
transport hierarchy and $I_c+1$ equations from the constraint hierarchy.
In order solve them the number of equations must at least 
equal the number of moments. However, if there are more equations than 
moments, the system may be overdetermined, and therefore we would like 
to require equality of the number of equations and moments:
 \begin{equation}
 \label{condition2}
   I_t+1 + I_c+1 = j_{\rm max}+1\, .
 \end{equation}
The two conditions (\ref{condition1}) and (\ref{condition2}) have a 
unique solution:
 \begin{equation}
 \label{solution}
   I_t = I_c+1 = M\, , \qquad j_{\rm max} = 2M\, ,
 \end{equation}
which yields $M+1$ transport and $M$ constraint equations for the first 
$2M+1$ energy moments. Smaller values of $j_{\rm max}$ don't yield enough 
equations, and larger values lead to an (at least superficially) 
overdetermined system of equations. For spinor fields ($M=0$) the 
truncated set involves only one transport and no constraint 
equation; for each spinor component it gives a single kinetic 
equation (the BGR equation \cite{BGR}) for its lowest energy moment, 
the equal-time Wigner function $W^s(x,\bbox{p})$. For scalar fields 
($M=1$) the truncated set contains two transport equations and one
constraint for the three lowest order moments $w_0,\, w_1, \, w_2$.

If we go beyond this minimal closed subset of moments and equations, 
we get two equations for every additional moment, one from the 
transport hierarchy and one from the constraint hierarchy. As we will 
show explicitly in the next Section, in the constraint hierarchy 
(\ref{hier2}) the highest moment always comes with a constant 
coefficient. As we increase the hierarchy index $i$ in 
Eqs.~(\ref{hier}), at each step the newly occurring moment can thus be 
explicitly expressed in terms of the already known lower order moments 
using the corresponding constraint equation from (\ref{hier2}). As we 
will discuss, these higher order constraint equations contain 
important physics.  But in addition, at each step there is also a 
dynamical equation of motion for the new moment from the transport 
hierarchy (\ref{hier1}). How can the two equations be consistent? The 
answer is that this transport equation is not an independent new 
equation, but (with some algebraic effort) can be expressed as a 
combination of the lower order equations which have already been used. 
Our proof of this fact uses explicitly the structure of the dynamical 
operators $\hat G_{mn}$ and $\hat F_{mn}$.  It involves cumbersome 
algebra, and only for scalar fields with only scalar potential or mean 
field interactions we have been able to find a general proof. For 
scalar and spinor QED the proof is still incomplete, and we will only 
demonstrate the first step for the $2M+2^{\rm nd}$ moment. A 
completion of the proof presumably requires a so far missing deeper 
insight into the general dynamic structure of the moment equations and 
their relation to the underlying covariant theory.  

\section{Scalar field theory}
\label{sec2c}

In this Section we will give an explicit and complete discussion of 
the moment hierarchy for the simplest case of a scalar field theory in 
Hartree approximation. We exemplify the truncation of the hierarchy 
and the recursive computation of the higher order moments beyond 
minimal truncation. The discussion in the following Section for the 
practically more relevant case of QED will be technically more 
involved and, unfortunately, also less complete.  

\subsection{Covariant kinetic equations}
\label{sec2c1}

Consider the Klein-Gordon equation with a scalar potential $U(x)$:
 \begin{equation}
 \label{klein}
  \left(\partial^2_x+m_0^2+U(x)\right)\hat\phi(x) = 0\ .
 \end{equation}
The covariant Wigner function is the four-dimensional Wigner transform 
of the covariant density matrix $\Phi(x,y) = \langle \hat \Phi(x,y) 
\rangle$: 
 \begin{equation}
    {\cal W}(x,p) = \int d^4y\, e^{ip{\cdot}y} \, \Phi(x,y)
    =\int d^4y\, e^{ip{\cdot}y}
     \left\langle \hat\phi \left(x+{\textstyle{y\over 2}}\right)
                  \hat\phi^+\left(x-{\textstyle{y\over 2}}\right)
     \right\rangle\, .
 \end{equation}
To derive the kinetic equations for the scalar Wigner function, we 
calculate the second-order derivatives $\left({1\over 2}\partial^x_\mu 
+\partial^y_\mu\right)^2$ and $\left({1\over 2}\partial^x_\mu -
\partial^y_\mu\right)^2$ of the covariant density operator, and then 
employ the Klein-Gordon equation (\ref{klein}) and its adjoint. After 
taking the ensemble average and performing the Wigner transform we 
obtain two complex kinetic equations: 
 \begin{mathletters}
 \label{complex}
 \begin{eqnarray}
 \label{complex1}
  \left({\textstyle{1\over 4}}\partial_x^2 - p^2 + m_0^2 
        + U\left(x-{\textstyle{i\over 2}}\partial_p\right)
        - i p{\cdot}\partial_x 
  \right) {\cal W}(x,p) &=& 0\, ,
 \\
 \label{complex2}
  \left({\textstyle{1\over 4}}\partial_x^2 - p^2 + m_0^2 
        + U\left(x+{\textstyle{i\over 2}}\partial_p\right)
        + i p{\cdot}\partial_x 
  \right) {\cal W}(x,p) &=& 0\, .
 \end{eqnarray}
 \end{mathletters}
Since the scalar Wigner function is real, adding and subtracting these 
two complex equations yields two real equations of the type 
(\ref{scalar}).  After reinstating $\hbar$ the corresponding operators 
$\hat G$ and $\hat F$ are given by 
 \begin{mathletters}
 \label{FG}
 \begin{eqnarray}
 \label{G}
  \hat G(x,p) &=& \hbar p{\cdot}\partial_x + {\rm Im\,} \hat M^2(x,p)\, ,
 \\
 \label{F}
  \hat F(x,p) &=& - p^2 + {\hbar^2\over 4}\partial_x^2 
                        + {\rm Re\,} \hat M^2(x,p)\, ,
 \end{eqnarray}
 \end{mathletters}
where the mass operator $\hat M^2$ is defined as 
 \begin{mathletters}
 \label{massop}
 \begin{eqnarray}
 \label{massop1}
   \hat M^2(x,p) &=& m_0^2+\hat\Sigma_e(x,p)+i\hat\Sigma_o(x,p)\, ,
 \\
 \label{massop2}
   \hat\Sigma_e(x,p) &=& \cos\left({\hbar\triangle\over 2}\right)U(x)\, ,
 \\
 \label{massop3}
   \hat\Sigma_o(x,p) &=& \sin\left({\hbar\triangle\over 2}\right)U(x)\, .
 \end{eqnarray}
 \end{mathletters}
Here the triangle operator $\triangle$ is defined as $\triangle = 
\partial_x{\cdot}\partial_p$ where the coordinate derivative 
$\partial_x$ acts only on the scalar potential $U(x)$ while 
$\partial_p$ acts only on the Wigner function.  

\subsection{Semiclassical expansion}
\label{sec2c2}

In the general quantum situation the particles have no definite mass 
due to quantum fluctuations around their classical mass shell and 
collision effects in the medium. In the situation here with only an 
external potential this is illustrated by the mass operator $\hat 
M^2$. Only in the classical limit $\hbar\to 0$ it reduces to the 
quasiparticle mass 
 \begin{mathletters}
 \label{quasi}
 \begin{eqnarray}
 \label{quasi1}
  {\rm Re\,} \hat M^2_0(x,p) &=& m^2(x) = m_0^2+U(x)\, ,
 \\
 \label{quasi2}
  {\rm Im\,} \hat M^2_0(x,p) &=& 0\ .
 \end{eqnarray}
 \end{mathletters}
In this case the constraint equation reduces to the on-shell condition
 \begin{equation}
 \label{ms1}
  \left(p^2-m^2(x)\right){\cal W}_0(x,p) = 0
 \end{equation}
for the classical covariant Wigner function ${\cal W}_0$. The 
classical transport equation arises from the general transport 
equation at first order in $\hbar$. The first order contribution to 
the mass operator is 
 \begin{eqnarray}
   {\rm Re\,} \hat M^2_1(x,p) &=& 0\, ,
 \nonumber\\
   {\rm Im\,} \hat M^2_1(x,p) &=& 
   \hbar\, m(x)\, (\partial_x m(x)){\cdot}\partial_p\, ,
 \end{eqnarray}
and we obtain the covariant Vlasov equation   
 \begin{equation}
  \Bigl(p{\cdot}\partial_x + m(x)(\partial_x m(x)){\cdot}\partial_p
  \Bigr) {\cal W}_0(x,p) = 0\, ,
 \end{equation}
with a Vlasov force term induced by $x$-dependent effective mass term.  
For scalar fields there is no first order quantum correction to the 
operator $F$ in (\ref{FG}), and from the zeroth order term we obtain 
the mass-shell condition for the first order Wigner function: 
 \begin{equation}
  \left(p^2-m^2(x)\right) {\cal W}_1(x,p) = 0\ .
 \end{equation}
This discussion holds universally for arbitrary potentials. If, for 
instance, $U(x)$ is generated by the scalar field $\hat\phi(x)$ itself 
in Hartree approximation, 
 \begin{equation}
  U(x) = -C+\lambda \langle\hat\phi(x)\hat\phi^+(x)\rangle
       = -C+\lambda \int{d^4 p\over (2\pi)^4}\, {\cal W}(x,p)\, ,
 \end{equation}
with a mass parameter $C$ and a coupling strength $\lambda$, this 
model provides a useful tool for a dynamical description of 
spontaneous symmetry breaking \cite{BGG}.  

\subsection{The 3-dimensional dynamical operators}
\label{sec2c3}
 
We now perform the energy average of the covariant transport and 
constraint equations (\ref{scalar}) and construct the hierarchy 
(\ref{hier}) of moment equations. The first step is the double 
expansion of the type (\ref{gradient}) for the covariant dynamical 
operators $\hat G(x,p)$ and $\hat F(x,p)$: 
 \begin{mathletters}
 \label{gfmn}
 \begin{eqnarray}
 \label{gmn}
  \hat G_{mn}(\bbox{x},\bbox{p},t) &=& \left\{\begin{array}{ll}
       \hbar \Lambda \partial_t & \mbox{for $m=1, n=0$}\\
       \hbar \bbox{p}{\cdot}\bbox{\nabla}_{\!x}-\hat\sigma_o & 
                         \mbox{for $m= n=0$}\\
       -{1\over n!}\left({i\hbar\over 2\Lambda}\right)^n
        \left( \partial_t^n \hat\sigma_o \right)& \mbox{for
                         $m=0, n\ne 0$ even}\\
       -{i\over n!}\left({i\hbar\over 2\Lambda}\right)^n
        \left( \partial_t^n \hat\sigma_e \right) & \mbox{for
                         $m=0, n$ odd}\\
       0  &\mbox{else}
       \end{array}\right.
 \\
 \label{fmn}
  \hat F_{mn}(\bbox{x},\bbox{p},t) &=& \left\{\begin{array}{ll}
       -\Lambda^2 & \mbox{for $m=2, n=0$}\\
       {\hbar^2\over 4}\partial_x^2 + \bbox{p}^2 + m_0^2 + \hat\sigma_e & 
                         \mbox{for $m=n=0$}\\
       {1\over n!}\left({i\hbar\over 2\Lambda}\right)^n
        \left(\partial_t^n \hat\sigma_e\right) & \mbox{for
                         $m=0, n\ne 0$ even}\\
       -{i\over n!}\left({i\hbar\over 2\Lambda}\right)^n
         \left(\partial_t^n \hat\sigma_o \right)& \mbox{for
                         $m=0, n$ odd}\\
       0  &\mbox{else.}
       \end{array}\right.
 \end{eqnarray}
 \end{mathletters}
Here
 \begin{mathletters}
 \label{sigma3}
 \begin{eqnarray}
 \label{sigma3e}
   \hat\sigma_e(\bbox{x},\bbox{p},t) &=& \cos\left({\hbar\over 2}
                         \bbox{\nabla}_{\!x}{\cdot}\bbox{\nabla}_{\!p}
                             \right)U(\bbox{x},t)\, ,
 \\
 \label{sigma3b}
   \hat\sigma_o(\bbox{x},\bbox{p},t) &=& \sin\left({\hbar\over 2}
                         \bbox{\nabla}_{\!x}{\cdot}\bbox{\nabla}_{\!p}
                             \right)U(\bbox{x},t)
 \end{eqnarray}
 \end{mathletters}
are the three-dimensional analogues of the covariant operators 
$\hat\Sigma_e$ and $\hat\Sigma_o$ in Eq.~(\ref{massop}). Again, the 
spatial gradients act only on $U(\bbox{x},t)$, while the momentum 
gradients act on the equal-time Wigner functions (i.e. on the energy 
moments $w_j(x,\bbox{p})$).  

The three-dimensional dynamical operators $\hat G_{mn}(x,\bbox{p})$ 
and $\hat F_{mn}(x,\bbox{p})$ must now be combined with the 
coefficients $c_{ij}^{mn}$ to obtain the dynamical operators $\hat 
g_{ij}(x,\bbox{p})$ and $\hat f_{ij}(x,\bbox{p})$ which are needed in 
the transport and constraint hierarchies. This is done in 
Appendix~\ref{appb}.  

\subsection{Minimal truncation}
\label{sec2c4}
 
The resulting transport hierarchy is truncated at $I_t=M=1$, the 
constraint hierarchy at $I_c=M-1=0$. This yields the following 
equations for $w_0(x,\bbox{p})$, $w_1(x,\bbox{p})$, and 
$w_2(x,\bbox{p})$: 
 \begin{mathletters}
 \label{mini}
 \begin{eqnarray}
 \label{mini1}
  && \hat g_{00}w_0+\hat g_{01}w_1 = 0\ ,\\
 \label{mini2}
  && \hat g_{10}w_0+\hat g_{11}w_1+\hat g_{12}w_2 = 0\ ,\\
 \label{mini3}
  && \hat f_{00}w_0+\hat f_{01}w_1+\hat f_{02}w_2 = 0\ .
 \end{eqnarray}
 \end{mathletters}
The dynamical operators $\hat g_{ij}$ and $\hat f_{ij}$ are given in 
Appendix~\ref{appb} and Eqs.~(\ref{gfmn}). Reexpressing $w_j$ in terms 
of the energy moments $W_j$ from Eq.~(\ref{moment1}), 
 \begin{mathletters}
 \label{reex}  
 \begin{eqnarray}
 \label{reex1}  
   w_0(x,\bbox{p}) &=& {1\over \Lambda \sqrt 2} W(\bbox{x},\bbox{p},t)\, ,
 \\
 \label{reex2}  
   w_1(x,\bbox{p}) &=& {1\over  \Lambda^2} \sqrt{3\over 2} 
                      W_1(\bbox{x},\bbox{p},t)\, ,
 \\
 \label{reex3}  
   w_2(x,\bbox{p}) &=& {1\over 2\Lambda}\sqrt{5\over 2}
                       \left({3\over \Lambda^2} W_2(\bbox{x},\bbox{p},t)
                             -W(\bbox{x},\bbox{p},t)\right)\, ,
 \end{eqnarray}
 \end{mathletters}
Eqs.~(\ref{mini}) can be rewritten as
 \begin{mathletters}
 \label{minimal}  
 \begin{eqnarray}
 \label{T1}  
   \partial_t W_1(\bbox{x},\bbox{p},t) 
   &=& -\left(\bbox{p}{\cdot}\bbox{\nabla}_{\!x} 
              - {1\over \hbar} \hat\sigma_o(\bbox{x},\bbox{p},t) \right)
        W(\bbox{x},\bbox{p},t)\, ,
 \\
 \label{T2}  
   \partial_t W_2(\bbox{x},\bbox{p},t) 
   &=& -\left(\bbox{p}{\cdot}\bbox{\nabla}_{\!x} 
              - {1\over \hbar} \hat\sigma_o(\bbox{x},\bbox{p},t) \right)
          W_1(\bbox{x},\bbox{p},t) 
       + {1\over 2} \left(\partial_t \hat\sigma_e(\bbox{x},\bbox{p},t)\right)
          W(\bbox{x},\bbox{p},t) \, , 
 \\
 \label{c1}
   W_2(\bbox{x},\bbox{p},t) 
   &=& \left( {\hbar^2 \over 4} 
              \left(\partial_t^2 - \bbox{\nabla}_{\!x}^2 \right) 
              + \bbox{p}^2 + m_0^2 + \hat\sigma_e(\bbox{x},\bbox{p},t) 
       \right) W(\bbox{x},\bbox{p},t) \, .
 \end{eqnarray}
 \end{mathletters}
Please note that all powers of the cutoff $\Lambda$ cancel in the 
final expressions as they should.  

The two transport equations (\ref{minimal}a,b) do not decouple, not 
even in the classical limit $\hbar \to 0$. To achieve decoupling one 
must return to the covariant equations in Sec.~\ref{sec2c1} and study 
their semiclassical limit as given in Sec.~\ref{sec2c2} {\em before} 
performing the energy average. Then the mass-shell condition 
(\ref{ms1}) can be used to rewrite all higher order energy moments in 
terms of the zeroth order moment as explained in the Introduction, 
Eq.~(\ref{class2}).  With this information the constraint (\ref{c1}), 
in the limit $\hbar \to 0$, becomes trivial, 
 \begin{equation}
 \label{miniclass1}
   W_2(\bbox{x},\bbox{p},t) = E_p^2(\bbox{x},t)\, W(\bbox{x},\bbox{p},t)\, ,
   \qquad   E_p^2(\bbox{x},t) = \bbox{p}^2 + m^2(\bbox{x},t)\, ,
 \end{equation}
while the two transport equations (\ref{T1}) and (\ref{T2}) become 
identical and can be written in the form of a Vlasov equation for the
charge density (see Sec.~\ref{sec3}):
 \begin{equation}
 \label{V1}  
   \partial_t W_1(\bbox{x},\bbox{p},t) 
   + \left({\bbox{p}\over E_p}{\cdot}\bbox{\nabla}_{\!x} 
              - \bbox{\nabla}_{\!x} E_p \cdot \bbox{\nabla}_{\!p} \right)
        W_1(\bbox{x},\bbox{p},t) = 0\, .
 \end{equation}
The reason why the information contained in Eq.~(\ref{class2}) cannot be 
easily recovered directly from the 3-dimensional transport and constraint 
equations is that in their derivation, through Eq.~(\ref{byparts}), we made 
heavy use of partial integration with respect to the energy. In the classical
limit this has the unfortunate effect of spreading the information 
contained in the on-shell condition over the whole infinite hierarchy of 
3-dimensional constraint equations.

Although we have always talked about Eq.~(\ref{c1}) as a ``constraint 
equation'' it is clear that, as far as solving the minimal subset 
(\ref{minimal}) of equal-time kinetic equations is concerned, 
this terminology is only adequate in the classical limit $\hbar \to 0$.
In general quantum situations it is a second order partial differential 
equation for the lowest order moment $W(\bbox{x},\bbox{p},t)$ which must 
be solved as an initial value problem together with the first order
partial differential equations (\ref{T1}) and (\ref{T2}).

\subsection{Higher order moment equations}
\label{sec2c5}

The next higher moment $w_3$ is determined by the third equation in 
the transport hierarchy and the second equation in the constraint 
hierarchy, 
 \begin{mathletters}
 \label{next}
 \begin{eqnarray}
 \label{next1}
  && \hat g_{20}w_0+\hat g_{21}w_1+\hat g_{22}w_2+\hat g_{23}w_3 = 0\ ,\\
 \label{next2}
  && \hat f_{10}w_0+\hat f_{11}w_1+\hat f_{12}w_2+\hat f_{13}w_3 = 0\ .
 \end{eqnarray}
 \end{mathletters}
With the dynamical operators from Appendix~\ref{appb} and 
Eqs.~(\ref{gfmn}) and 
 \begin{equation}
 \label{reex4}  
   w_3(x,\bbox{p}) = {1\over 2\Lambda}\sqrt{7\over 2}
                  \left({5\over \Lambda^3} W_3(\bbox{x},\bbox{p},t)
                        - {3\over \Lambda} W_1(\bbox{x},\bbox{p},t)\right)
 \end{equation}
we obtain (using Eq.~(\ref{T1}))
 \begin{mathletters}
 \label{explizit}  
 \begin{eqnarray}
 \label{T3}  
   \partial_t W_3(\bbox{x},\bbox{p},t) 
   &=& -\left(\bbox{p}{\cdot}\bbox{\nabla}_{\!x} 
              - {1\over \hbar} \hat\sigma_o(\bbox{x},\bbox{p},t) \right)
          W_2(\bbox{x},\bbox{p},t) 
       +  \left(\partial_t \hat\sigma_e(\bbox{x},\bbox{p},t)\right)
          W_1(\bbox{x},\bbox{p},t) 
 \nonumber\\
   &&  -  {\hbar\over 4} \left(\partial_t^2 
                         \hat\sigma_o(\bbox{x},\bbox{p},t)\right)
          W(\bbox{x},\bbox{p},t) \, , 
 \\
 \label{c2}  
   W_3(\bbox{x},\bbox{p},t) 
   &=& \left( {\hbar^2 \over 4} 
              \left(\partial_t^2 - \bbox{\nabla}_{\!x}^2 \right) 
              + \bbox{p}^2 + m_0^2 + \hat\sigma_e(\bbox{x},\bbox{p},t) 
       \right) W_1(\bbox{x},\bbox{p},t) 
 \nonumber\\
   &&  -  {\hbar\over 2} \left(\partial_t 
                         \hat\sigma_o(\bbox{x},\bbox{p},t)\right)
          W(\bbox{x},\bbox{p},t) \, . 
 \end{eqnarray}
 \end{mathletters}
By substituting (\ref{next2}) into (\ref{next1}) and taking into account
the following commutators:
 \begin{mathletters}
 \label{commutator}  
 \begin{eqnarray}
 \label{com1}
  && [\hat G_{10},\ \hat F_{00}] = - 2 \hat G_{01} \hat F_{20}\ ,
 \\ 
 \label{com2}
  && [\hat G_{00},\ \hat F_{00}] = \hat F_{01} \hat G_{10} 
     - 2 \hat G_{02} \hat F_{20} \ ,
 \\
 \label{com3}
  && [{\hbar^2\over 4}\partial^2,\ \hat\sigma_{e/o}] 
     = {\hbar^2\over 2} \partial\hat\sigma_{e/o}{\cdot}\partial
     + {\hbar^2\over 4}\partial^2 \hat\sigma_{e/o}\ ,
 \\
 \label{com4}
  && [\bbox{p}{\cdot}\bbox{\nabla}_{\!x},\ \hat\sigma_{e/o}] 
     = \bbox{p}{\cdot}\bbox{\nabla}_{\!x}\hat\sigma_{e/o}
     + {\hbar\over 2}\bbox{\nabla}_{\!x}
       \hat\sigma_{o/e}{\cdot}\bbox{\nabla}_{\!x}\ ,
 \\
 \label{com5}
  && [\bbox{p}^2,\ \hat\sigma_{e/o}] 
     = -\hbar \bbox{p}{\cdot}\bbox{\nabla}_{\!x} \hat\sigma_{o/e}
     + {\hbar^2\over 4}\bbox{\nabla}_{\!x}^2\hat\sigma_{e/o}\ ,
 \end{eqnarray}
 \end{mathletters}
the third transport equation (\ref{next1}) can be rewritten in terms of
the first transport equation (\ref{mini1}) as
 \begin{equation}
 \label{w30}
  \hat f_{00}\left(\hat g_{00}w_0+\hat g_{01}w_1\right) = 0\, .
 \end{equation}
This implies that the transport equation (\ref{next1}) for $w_3$ is 
redundant. The third-order moment $W_3$ is completely determined in 
terms of the solutions of the minimal subgroup (\ref{minimal}) by the 
constraint equation (\ref{c2}). It arises from Eq.~(\ref{next2}) by 
noting that $\hat f_{13}$ is a constant, $\hat f_{13} = C_3 = 
-{2\sqrt{3}\over 5 \sqrt{7}}\Lambda^2$ (see Appendix~\ref{appb}), and 
solving for $w_3$: 
 \begin{equation}
 \label{w3}
  w_3 = -{1\over C_3}\left(\hat f_{10} w_0 + \hat f_{11}w_1 
                         + \hat f_{12} w_2 \right) \, .
 \end{equation}
Note that, in contrast to (\ref{c1}), Eq.~(\ref{c2}) does not require 
solving a partial differential equation because everything on the r.h.s. 
is known from the solution of Eqs.~(\ref{minimal}).

The above procedure can be extended to all higher order moment 
equations, by repeatedly using the commutators listed in 
(\ref{commutator}). In general one finds that a transport equation 
 \begin{equation}
  \sum_{j=0}^{i+1}\hat g_{ij}w_j = 0
 \end{equation}
with $i\geq 2$ can be re-expressed in terms of the first $(i-1)$ 
transport equations as 
 \begin{equation}
  \sum_{j=0}^{i-2}\hat f_{i-2,j} \left(\sum_{k=0}^{j+1}
  \hat g_{jk}w_k\right) = 0\, .
 \end{equation}
Thus, except for the first two, all transport equations are redundant.  
The higher order moments $w_i$ with $i>2$ can be computed from their 
constraint equations 
 \begin{equation}
 \label{recu}
  w_i = -{1\over C_i} \sum_{j=0}^{i-1}\hat f_{i-2,j}w_j\, ,
 \end{equation}
with the constant $C_i$ given by
 \begin{equation}
  C_i\equiv \hat f_{i-2,i} = -{i(i-1)\over (2i-1)\sqrt{(2i-3)(2i+1)}}\ .
 \end{equation}


\section{Application to QED}
\label{sec3}

\subsection{Scalar QED}
\label{sec3a}

In this Section we discuss the application of the general formalism 
developed in Sec.~\ref{sec2} to QED. Since some of the equations will 
be rather lengthy we will economize on the notation by dropping all 
factors of $\hbar$. The latter are correctly given in 
Refs.~\cite{ZH1,ZH2} to which we refer in case of need.  

We begin with the case of scalar QED with external electromagnetic 
fields. In Ref.~\cite{ZH1} we discussed the semiclassical transport 
equations for this theory by energy averaging the semiclassical limit 
of the covariant transport equations. In this subsection we will 
derive the general equal-time quantum transport equations by 
performing the energy average without any approximations. In the 
following subsection the result will be compared with the 
corresponding equations derived by directly Wigner-transforming the 
equations of motion for the equal-time density matrix \cite{BGG}.  

In scalar QED the scalar field obeys the Klein-Gordon equation
 \begin{equation}
 \label{kg} 
   \left( (\partial_\mu + ieA_\mu(x)) (\partial^\mu+ieA^\mu(x))
     + m^2 \right) \, \phi(x) = 0 \ ,
 \end{equation}
Following an analogous procedure as in Sec.~\ref{sec2c1} (see 
\cite{ZH1} for details) one derives two covariant kinetic equations of 
type (\ref{scalar}) for the covariant Wigner function 
 \begin{equation}
    {\cal W}(x,p) = \int d^4y\, e^{ip{\cdot}y}
     \left\langle \hat\phi \left(x+{\textstyle{y\over 2}}\right)
     \exp\left[ ie \int_{-1/2}^{1/2} ds\, A(x+sy){\cdot}y \right]
                  \hat\phi^+\left(x-{\textstyle{y\over 2}}\right)
     \right\rangle\, .
 \end{equation}
The corresponding covariant dynamical operators $\hat G$ and $\hat F$ 
are given by 
 \begin{mathletters}
 \label{gf}
 \begin{eqnarray}
 \label{gf1}
   \hat G(x,p) &=& \hat\Pi^\mu(x,p)\hat D_\mu(x,p)\,,
 \\
 \label{gf2}
   \hat F(x,p) &=& {1\over 4}\hat D^\mu(x,p)\hat D_\mu(x,p)
                  -\hat\Pi^\mu(x,p) \hat\Pi_\mu(x,p)+m^2\,,
 \\
 \label{gf3}
   \hat\Pi_\mu(x,p)&=& p_\mu - ie\int ^{1\over 2}_{-{1\over 2}} ds\, s\, 
                   F_{\mu\nu}(x-is\partial_p) \, \partial_p^\nu \,,
 \\
 \label{gf4}
   \hat D_\mu(x,p)  &=& \partial_\mu - e\int ^{1\over 2}_
                        {-{1\over 2}} ds\,
                        F_{\mu\nu} (x-is\partial_p)\, \partial_p^\nu \,.
 \end{eqnarray}
 \end{mathletters}
$F^{\mu\nu}=\partial^\mu A^\nu-\partial^\nu A^\mu$ is the 
electromagnetic field tensor.  

The structure of the equal-time transport theory for scalar QED is 
very similar to that for a scalar potential $U(x)$ which we considered 
in the previous section. The difference resides solely in the 
expressions for the dynamical operators $\hat G_{mn}$ and $\hat 
F_{mn}$. In Appendix~\ref{appc} we provide the double expansions 
(\ref{gradient}) for the basic operators $\hat\Pi^\mu$ and $\hat 
D^\mu$ as well as for $\hat G$ and $\hat F$.  

In terms of the covariant Wigner function the charge current $j^\mu$ 
and the energy-momentum tensor $T^{\mu\nu}$ are expressed as 
 \begin{mathletters}
 \label{jt}
 \begin{eqnarray}
 \label{jmu}
   &&j^\mu(x,p) = e\, p^\mu\, {\cal W}(x,p)\,,
 \\
  \label{tmunu}
   &&T^{\mu\nu}(x,p) = p^\mu\, p^\nu\, {\cal W}(x,p)\,.
 \end{eqnarray}
 \end{mathletters}
After performing the energy average these equations translate into 
relations between the first three moments $w_0,w_1,w_2$ and the 
equal-time phase-space distributions for the scalar density 
$W(\bbox{x},\bbox{p},t)$, the charge density 
$\rho(\bbox{x},\bbox{p},t)$, and the energy density 
$\epsilon(\bbox{x},\bbox{p},t)$: 
 \begin{mathletters}
 \label{relation}
 \begin{eqnarray}
 \label{relation1}  
   w_0(\bbox{x},\bbox{p},t) 
   &=& {1\over \Lambda \sqrt 2} W(\bbox{x},\bbox{p},t)\ ,
 \\
 \label{relation2}  
   w_1(\bbox{x},\bbox{p},t) 
   &=& \sqrt{3\over 2} {1\over e\, \Lambda^2} \rho(\bbox{x},\bbox{p},t)\ ,
 \\
 \label{relation3}  
   w_2(\bbox{x},\bbox{p},t) &=& {1\over 2\Lambda}\sqrt{5\over 2}
                  \left({3\over \Lambda^2}\epsilon(\bbox{x},\bbox{p},t)
                        -W(\bbox{x},\bbox{p},t)\right)\ .
 \end{eqnarray}
 \end{mathletters}
The charge current density $\bbox{j}(\bbox{x},\bbox{p},t)$ and the 
momentum density $\bbox{P}(\bbox{x},\bbox{p},t)$ can be expressed in 
terms of $W$ and $\rho$ as 
 \begin{equation}
 \label{cm}
   \bbox{j}(\bbox{x},\bbox{p},t) = e\,\bbox{p}\,W(\bbox{x},\bbox{p},t)\, ,
 \qquad
   \bbox{P}(\bbox{x},\bbox{p},t) = 
   {\bbox{p}\over e}\,\rho(\bbox{x},\bbox{p},t)\,.
 \end{equation}
The subgroup (\ref{mini}) determining the moments $w_0$, $w_1$ and 
$w_2$ can thus be equivalently rewritten as transport and constraint 
equations for $W$, $\rho$ and $\epsilon$: 
 \begin{mathletters}
 \label{wce}
 \begin{eqnarray}
 \label{wce1}
   && {1\over e}\hat d_t \rho +(\hat d_t\hat \pi_0
       +\hat{\bbox{\pi}}{\cdot}{\hat{\bbox{d}}}) W = 0\ ,
 \\
 \label{wce2}
   && \hat d_t\epsilon +{1\over e}(\hat d_t\hat \pi_0
     +{\hat{\bbox{\pi}}}{\cdot}{\hat{\bbox{d}}}+\hat D)\rho
     +(\hat \pi_0\hat D+{\hat{\bbox{\pi}}}{\cdot}{\hat{\bbox{I}}}
     -\hat d_t \hat A
     +{\hat{\bbox{G}}}{\cdot}{\hat{\bbox{d}}})W=0\ ,
 \\
 \label{wce3}
   && \epsilon = \left({1\over 4} (\hat d_t^2-{\hat{\bbox{d}}}^2)
      -(\hat\pi_0^2 -{\hat{\bbox{\pi}}}^2)+m^2+\hat A\right)W 
      -{2\hat\pi_0\over e}\rho\ ,
 \end{eqnarray}
 \end{mathletters}
The expressions of $\hat g_{ij}$ and $\hat f_{ij}$ were obtained via 
the relations given in Appendix~\ref{appb} from the equal-time 
operators (\ref{C2}) in Appendix~\ref{appc}. We have also used the 
commutators 
 \begin{mathletters}
 \label{comm1}
 \begin{eqnarray}
 \label{comm1a}
    [\hat d_t,\ \hat\pi_0] &=& \hat D\ ,
 \\
 \label{comm1b}
    [\hat d_t,\ \hat A]    &=& 2\hat E\ .
 \end{eqnarray}
 \end{mathletters}
Due to the line integrals over $s$ in the operators $\hat\Pi_\mu$ and 
$\hat D_\mu$ which guarantee the gauge invariance of the formalism 
\cite{EH}, the equal-time operators $\hat G_{mn}$ and $\hat F_{mn}$ 
for QED are much more complicated than Eqs.~(\ref{gfmn}) for the case 
of a scalar potential. This is the origin of the more complicated 
structure of the minimal subgroup (\ref{wce}) of 3-dimensional kinetic 
equations. Please note the sequence of the operators in 
Eqs.~(\ref{wce}): in particular the generalized time derivatives $\hat 
d_t$ act on everything following them. Eqs.~(\ref{wce}) are thus much 
more intricately coupled than the corresponding equations 
(\ref{minimal}) for the pure scalar case.  

It is instructive to integrate Eqs.~(\ref{wce}) over $\bbox{p}$ to 
obtain equations of motion for the corresponding space-time densities: 
 \begin{mathletters}
 \label{conslaw}
 \begin{eqnarray}
 \label{currcons}
   &&\partial_t \rho(\bbox{x},t) 
     + \bbox{\nabla}{\cdot}\bbox{j}(\bbox{x},t)= 0\, ,
 \\
 \label{encons}
   &&\partial_t \epsilon(\bbox{x},t)
     + \bbox{\nabla}{\cdot}\bbox{P}(\bbox{x},t)
     - \bbox{E}(\bbox{x},t){\cdot}\bbox{j}(\bbox{x},t) = 0\, ,
 \\
 \label{endens}
   &&\epsilon(\bbox{x},t) = \int{d^3 p\over (2\pi)^3} \left({\hbar^2\over 4}
                 (\partial_t^2-\bbox{\nabla}_{\!x}^2)+\bbox{p}^2
                 +m^2\right)W(\bbox{x},\bbox{p},t) \, .
 \end{eqnarray}
 \end{mathletters}
The first two equations express the conservation of electric charge 
and of energy-momentum while the last equation gives the energy 
density in terms of the scalar equal-time phase-space density 
$W(\bbox{x},\bbox{p},t)$ including quantum corrections. In 
Eq.~(\ref{encons}) $\bbox{P}(\bbox{x},t) = \int (d^3p/(2\pi)^3) 
\bbox{P}(\bbox{x},\bbox{p},t)$ is the momentum density in coordinate 
space, and the last term describes the conversion of field energy into 
mechanical energy by the work done by the electric field on moving 
charges.  
   
Let us now consider the next two equations in the hierarchy, the 
transport and constraint equations (\ref{next}) for the third-order 
moment $w_3$. Again the constraint equation can be directly solved for 
$w_3$ in terms of the solutions $w_0,w_1,w_2$ of 
Eqs.~(\ref{relation}), (\ref{wce}) (see Eq.~(\ref{w3})). Our task is 
to show that the transport equation for $w_3$ is redundant. To this 
end we substitute the constraint (\ref{next2}) into the transport 
equation (\ref{next1}) and use the scalar QED analogue of the 
commutators (\ref{commutator}), namely (\ref{comm1}) and 
 \begin{mathletters}
 \label{comm2}
 \begin{eqnarray}
 \label{comm2a}
    && [\hat d_t,\ \hat B] = 3 \hat F\, ,
 \\
 \label{comm2b}
    && [\hat d_t,\ {\hat{\bbox{\pi}}}] = -\bbox{I}
       +\bbox{\nabla}_{\!x}\hat\pi_0\, ,
 \\
 \label{comm2c}
    && [{\hat{\bbox{I}}}, \ {\hat{\bbox{\pi}}}] 
       = -{\bbox{\nabla}}_{\!x}{\cdot}{\hat{\bbox{G}}}\, ,
 \\
 \label{comm2d}
    && [{\hat{\bbox{d}}}, \ \hat\pi_0] = {\bbox{\nabla}}_{\!x} \hat\pi_0\, ,
 \\
 \label{comm2e}
    && [\hat d_t,\ {\hat{\bbox{d}}}] 
       = -e{\bbox{\nabla}_{\!x}}{\cdot}{\bbox{\nabla}}_{\!p}
       \int^{1\over 2}_{-{1\over 2}} ds \, 
       {\bbox{E}}(\bbox{x}+is{\bbox{\nabla}}_{\!p}, t)\, ,
 \\
 \label{comm2f}
    && [{\hat{\bbox{\pi}}},\ \hat\pi_0] = -ie\int^{1\over 2}_{-{1\over 2}}
       ds\, s\, \bbox{E}(\bbox{x}+is\bbox{\nabla}_{\!p},t)
       + e\bbox{\nabla}_{\!x} \int^{1\over 2}_{-{1\over 2}}ds\, s^2\,
       \bbox{E}(\bbox{x}+is\bbox{\nabla}_{\!p},t){\cdot}\bbox{\nabla}_{\!p}\, ,
 \end{eqnarray}
 \end{mathletters}
as well as the identity
 \begin{equation}
 \label{int}
   \int^{1\over 2}_{-{1\over 2}}ds \left(2is+{1\over 4}
   \bbox{\nabla}_{\!x}{\cdot}\bbox{\nabla}_{\!p}
   -s^2\bbox{\nabla}_{\!x}{\cdot}\bbox{\nabla}_{\!p}\right)
    \bbox{E}(\bbox{x}+is\bbox{\nabla}_{\!p},t) = 0\ ,
 \end{equation}
to rewrite (\ref{next1}) in the form
 \begin{equation}
 \label{next3}
   \hat f_{00}\left(\hat g_{00}w_0+\hat g_{01}w_1\right)+\hat f_{01}
   \left(\hat g_{10}w_0 +\hat g_{11}w_1 +\hat g_{12}w_2\right) = 0\ .
 \end{equation}
Note that in the derivation of the commutators (\ref{comm2}) we used 
Maxwell's equation
 \begin{equation}
    \partial_\mu \tilde F^{\mu\nu} = 0\ ,
 \end{equation}
for the dual field strength tensor $\tilde F_{\mu\nu} = {1\over 2} 
\epsilon_{\mu\nu\sigma\rho} F^{\sigma\rho}$. The identity (\ref{int}) 
(which has no analogue in a theory without gauge invariance) is proven 
by expanding the electric field $\bbox{E}(\bbox{x} + i s 
\bbox{\nabla}_{\!p}, t) = e^{is\bbox{\nabla}_{\!x}{\cdot}\bbox{\nabla}_{\!p}} 
\bbox{E}(\bbox{x},t)$ and integrating term by term. Eq.~(\ref{next3}) 
expresses the transport equation for $w_3$ in terms of the transport 
equations for $w_1$ and $w_2$. This proves that it is redundant.  

>From the comparison of (\ref{next3}) with (\ref{w30}) we observe that 
in the case of scalar QED the third transport equation is expressed in 
terms of the first {\em and} second transport equations, while only 
the first one occurs in the case of scalar potentials. This difference 
results from the line integrals which occur in the gauge theory. In a 
transport theory with gauge invariance, the kinetic momentum is not 
$p_\mu$ but $\hat \Pi_\mu$. The energy average of the zeroth component 
of the second term in (\ref{gf3}) yields the equal-time operator $\hat 
\pi_0$ given in (\ref{C2}), and $\hat \pi_0= - {\sqrt{3}\over 
2\Lambda} \hat f_{01}$ in turn generates the coefficient in front of 
the second bracket in (\ref{next3}). In a transport theory without 
gauge freedom all coefficients $\hat f_{i,i+1}$ vanish due to the 
absence of linear terms in $p_\mu$ in the covariant constraint 
equation.  
 
We have not had the patience to carry the above considerations to 
higher orders in the energy moments. The corresponding calculations 
become extremely messy. Based on the experience with pure scalar 
theory we expect all higher order transport equations to be redundant, 
but we have failed to discover a simple calculational technique which 
permits us to prove this in an elegant way.  
   
\subsection{Comparison with the Feshbach-Villars approach}
\label{sec3b}

In Ref.~\cite{BGG} a set of equal-time transport equations for scalar 
QED was derived by directly Wigner transforming the equations of 
motion for the (non-covariant) equal-time density matrix. Since the 
Klein-Gordon equation (\ref{kg}) contains a second-order time 
derivative, its direct translation into 3-dimensional phase space does 
not lead to a sensible transport equation which should have only 
first-order time derivatives. Therefore the 3-dimensional approach 
exploits the Feshbach-Villars representation \cite{FV} of the field 
equations of motion which contains only first-order time derivatives. 
The price to pay (in addition to the loss of manifest Lorentz 
covariance) is the introduction of an auxiliary field which results in 
a rather complicated $2\times 2$ matrix structure of the scalar Wigner 
function. For the energy averaging method advocated here there is no 
such problem. In this case second-order time derivatives appear only 
in the constraint hierarchy, and the transport hierarchy contains only 
first-order time derivatives.  
  
Expressing the equal-time density operator $\Phi(\bbox{x},\bbox{y},t)$ 
of the scalar field in terms of the Feshbach-Villars spinors, 
performing a 3-dimensional Wigner transformation and using the 
Klein-Gordon equation in Feshbach-Villars form, the authors of 
\cite{BGG} obtained the following coupled equations for the $4$ spinor 
components $f_i(\bbox{x},\bbox{p},t)\ (i=0,1,2,3)$ of the Wigner 
function: 
 \begin{mathletters}
 \label{f0123}
 \begin{eqnarray}
 \label{f0}
     m \hat d_t f_0 
     &=& -{\hat{\bbox{\pi}}}{\cdot}{\hat{\bbox{d}}}(f_2+f_3)\, ,
 \\
 \label{f1}
     m \hat d_t f_1 &=& -({1\over 4}{\hat{\bbox{d}}}^2
     -{\hat{\bbox{\pi}}}^2)(f_2+f_3) + 2m^2 f_2 \, ,
 \\
 \label{f2}
     m \hat d_t f_2 &=& ({1\over 4}{\hat{\bbox{d}}}^2-{\hat{\bbox{\pi}}}^2)
        f_1 + {\hat{\bbox{\pi}}}{\cdot}{\hat{\bbox{d}}} f_0 - 2m^2 f_1 \, ,
 \\
 \label{f3}
     m \hat d_t f_3 &=& -({1\over 4}{\hat{\bbox{d}}}^2
     - {\hat{\bbox{\pi}}}^2) f_1 
     - {\hat{\bbox{\pi}}}{\cdot}{\hat{\bbox{d}}} f_0 \, . 
 \end{eqnarray}
 \end{mathletters}
In Ref.~\cite{BGG} these equal-time phase-space distributions were 
related to the physical phase-space densities $\rho,\, \bbox{j}, 
\epsilon$, and $\bbox{P}$ as follows \cite{BGG,Best}: 
 \begin{mathletters}
 \label{density}
 \begin{eqnarray}
 \label{density1}
    \rho &=& e f_0\, ,
 \\
 \label{density2}
    \epsilon &=& {\bbox{p}^2\over 2m}\Bigl(f_2+f_3\Bigr)+m f_3\, ,
 \\
 \label{density3}
    \bbox{j} &=& {e\over m} \bbox{p}\Bigl(f_2+f_3\Bigr)\, ,
 \\
 \label{density4}
    \bbox{P} &=& \bbox{p} f_0\, .
 \end{eqnarray}
 \end{mathletters}
Unlike the spinor decomposition for spinor QED \cite{VGE,BGR} where 
each component of the Wigner function has a definite physical meaning, 
there is no obvious physical interpretation for the component $f_1$ in 
scalar QED.  

The comparison of the charge currents $\bbox{j}$ in 
Eqs.~(\ref{density3}) and (\ref{cm}) leads directly to the 
identification 
 \begin{equation}
 \label{f23}
   W = {1\over m} \Bigl( f_2+f_3\Bigr)\, .
 \end{equation}
Using the relations (\ref{density}) and (\ref{f23}) one finds from a 
suitable linear combination of the kinetic equations (\ref{f0123}) the 
following equations of motion for the physical phase-space 
distributions $\rho$ and $\epsilon$, 
 \begin{mathletters}
 \label{FV}
 \begin{eqnarray}
 \label{FV1}
   &&{1\over e}\hat d_t \rho 
   + {\hat{\bbox{\pi}}}{\cdot}{\hat{\bbox{d}}} W=0\, ,
 \\
 \label{FV2}
   &&\hat d_t \epsilon
     +{1\over e}{\hat{\bbox{\pi}}}{\cdot}{\hat{\bbox{d}}}\rho
     -{1\over 2}\left(\Bigl({{\hat{\bbox{d}}}^2\over 4}
                            -{\hat{\bbox{\pi}}}^2\Bigr)\hat d_t
                      +\hat d_t \bbox{p}^2\right)W=0\, ,
 \\
 \label{FV3}
   &&\epsilon = \left({1\over 4}\Bigl(\hat d_t^2
                                      -{1\over 2}{\hat{\bbox{d}}}^2\Bigr)
                +{1\over 2}({\hat{\bbox{\pi}}}^2+\bbox{p}^2)+m^2\right)W\, ,
 \end{eqnarray}
 \end{mathletters}
as well as the following relation between the unphysical component 
$f_1$ and the scalar density: 
 \begin{equation}
 \label{FV4}
   f_1 = -{1\over 2}\hat d_t W\ .
 \end{equation}

Comparing Eqs.~(\ref{FV}) from the equal-time Feshbach-Villars 
approach with the corresponding Eqs.~(\ref{wce}) from the covariant 
approach using our energy-averaging procedure, one realizes that they 
have a very different structure. The difference is even more obvious 
when looking at the corresponding equations for the space-time 
densities which are obtained by integrating Eqs.~(\ref{FV}) over 
momentum: 
 \begin{mathletters}
 \label{conslaw1}
 \begin{eqnarray}
 \label{currcons1}
   &&\partial_t \rho(\bbox{x},t) 
     + \bbox{\nabla}{\cdot}\bbox{j}(\bbox{x},t)= 0\, ,
 \\
 \label{encons1}
   &&\partial_t \epsilon(\bbox{x},t) 
     + \bbox{\nabla}{\cdot}\bbox{P}(\bbox{x},t)
     - \bbox{E}(\bbox{x},t){\cdot}\bbox{j}(\bbox{x},t) 
     - {\hbar^2 \nabla_{\!x}^2 \over 8} \partial_t 
       \int{d^3p\over (2\pi)^3} W(\bbox{x},\bbox{p},t) = 0\, ,
 \\
 \label{endens1}
   &&\epsilon(x) = \int{d^3p\over (2\pi)^3} \left({1\over 4}
            \Bigl(\partial_t^2-{1\over 2}\bbox{\nabla}_{\!x}^2\Bigr)
            +\bbox{p}^2+m^2\right) W(\bbox{x},\bbox{p},t) \, .
 \end{eqnarray}
 \end{mathletters}
The first equation describes charge conservation and agrees with our 
Eq.~(\ref{currcons}). When comparing Eqs.~(\ref{encons}) and 
(\ref{encons1}) one sees that the energy-momentum conservation law 
derived from the Feshbach-Villars approach features a strange 
additional term of unknown physical origin. Similarly, when compared 
to Eq.~(\ref{endens}), Eq.~(\ref{endens1}) contains a strange factor 
${1\over 2}$ in front of the Laplace operator which breaks Lorentz 
invariance. We believe that these discrepancies indicate a technical 
problem with the Feshbach-Villars approach of Ref.~\cite{BGG}. Whether 
the problem resides in the derivation of the transport equations 
(\ref{f0123}) or in the postulated relations (\ref{density}) between 
the Feshbach-Villars spinor components $f_i$ and the physical 
densities could not be clarified. Only in the case of spatially 
constant electromagnetic fields (which to our knowledge is the only 
situation to which these equations have so far been applied 
\cite{BGG,ZH1}) the two sets of equations (\ref{conslaw}) and 
(\ref{conslaw1}) coincide. If we further neglect the magnetic field 
$\bbox{B}$, we derive two decoupled ordinary differential equations in 
time for $W$ and $\rho$: 
 \begin{eqnarray}
 \label{constant}
   &&d_t \rho =0\ ,\nonumber\\
   &&\Bigl( d_t^3+4(\bbox{p}^2+m^2)d_t
           +4e\bbox{E}{\cdot}\bbox{p}\Bigr) W = 0\ .
 \end{eqnarray}
These equations are identical in both approaches. The first equation 
expresses charge conservation in a homogeneous electric field, and the 
second equation is the well-known equation of motion for the charge 
current $\bbox{j}$ \cite{BE,ZH1} which, according to the relation 
$\bbox{j} = e{\bbox{p}\over \sqrt{\bbox{p}^2+m^2}}n$, governs the time 
evolution of the particle density $n$ by pair production in the 
electric field. This latter quantity thus comes out the same in both 
approaches.  

\subsection{Spinor QED}
\label{sec3c}

The case of spinor QED has been discussed in the context of the energy 
averaging method in Refs.~\cite{ZH1,ZH2,ABZH}. The discussion 
presented in those papers has, however, focussed entirely on the 
equal-time kinetic equations for the lowest energy moment of the 
covariant Wigner function. Here we will reformulate the problem in 
terms of the equal-time hierarchy of energy moments as introduced in 
Sec.~\ref{sec2} and also discuss the equations of motion for the 
higher order moments.  

Let us briefly review the relevant technical steps. We begin by 
performing a spinor decomposition \cite{VGE} of the covariant Wigner 
function, separating in a second step explicitly the temporal and 
spatial parts \cite{BGR} of the covariant spinor components: 
 \begin{mathletters}
 \label{spindec}
 \begin{eqnarray}
 \label{spindec1}
    W(x,p) &=& {1\over 4}\left[F(x,p)+i\gamma_5P(x,p)
               +\gamma_\mu V^\mu(x,p)
               +\gamma_\mu\gamma_5A^\mu(x,p)
               +{1\over 2}\sigma_{\mu\nu}S^{\mu\nu}(x,p)\right]
 \\
 \label{spindec2}
        &=& {1\over 4}\Bigl[
               \gamma_0 F_0(x,p) + \gamma_5 \gamma_0 F_1(x,p)
            + i\gamma_5 F_2(x,p) + F_3(x,p) 
 \nonumber\\
        & & -\gamma_5 \bbox{\gamma}{\cdot}\bbox{G}_0(x,p)
            -\bbox{\gamma}{\cdot}\bbox{G}_1(x,p)
            +i\gamma_5 \bbox{\gamma}{\cdot}\bbox{G}_2(x,p)
            +\gamma_5 \gamma_0 \bbox{\gamma}{\cdot}\bbox{G}_3(x,p)
            \Bigr]\ .
 \end{eqnarray}
 \end{mathletters}
Inserting the decomposition (\ref{spindec1}) into the covariant (VGE) 
equation of motion \cite{VGE} for the Wigner function $W(x,p)$, 
 \begin{equation}
 \label{vge}
   \left(\gamma^\mu(\hat \Pi_\mu(x,p)+{i\over 2}\hat D_\mu(x,p))
   -m\right) W(x,p) = 0\ , 
 \end{equation}
with $\hat \Pi_\mu(x,p)$ and $\hat D_\mu(x,p)$ from 
Eqs.~(\ref{gf}c,d), and separating real and imaginary parts one 
arrives at two groups of coupled covariant kinetic equations
of type (\ref{spinor}), see Eqs.~(74,75) of Ref.~\cite{ZH1}.  
Performing the energy average then leads to two groups of equal-time 
kinetic hierarchies (with 16 such hierarchies of equations in each 
group) for the energy moments of the 16 covariant spinor components. 

Since for spinor QED $M=0$, minimal truncation of these $2\times16$
hierarchies according to Sec.~\ref{sec2b} results in {\em one} transport 
equation from each of the 16 transport hierarchies and {\em no} constraint
equations. The minimal subgroup of equal-time kinetic equations thus
consists only of the 16 transport equations for the 16 zeroth-order
energy moments $f_i(\bbox{x},\bbox{p},t)$ and $\bbox{g}_i(\bbox{x},
\bbox{p},t)$ of the covariant spinor components $F_i(x,p)$ 
and $\bbox{G}_i(x,p)$ ($i=0,1,2,3$): 
 \begin{mathletters}
 \label{bgr}
 \begin{eqnarray}
 \label{bgr1}
  && \hat d_t f_0+{\hat{\bbox{d}}}{\cdot}\bbox{g}_1 = 0\ ,
 \\
 \label{bgr2}
  && \hat d_t f_1+{\hat{\bbox{d}}}{\cdot}\bbox{g}_0 +2 m f_2 = 0\ ,
 \\
 \label{bgr3}
  && \hat d_t f_2+2{\hat{\bbox{\pi}}}{\cdot}\bbox{g}_3 - 2 m f_1 = 0\ ,
 \\
 \label{bgr4}
  && \hat d_t f_3-2{\hat{\bbox{\pi}}}{\cdot}\bbox{g}_2 = 0\ ,
 \\
 \label{bgr5}
  && \hat d_t \bbox{g}_0+{\hat{\bbox{d}}}f_1
    -2{\hat{\bbox{\pi}}}\times\bbox{g}_1 = 0\ ,
 \\
 \label{bgr6}
  && \hat d_t \bbox{g}_1+{\hat{\bbox{d}}}f_0
     -2{\hat{\bbox{\pi}}}\times\bbox{g}_0 + 2 m \bbox{g}_2 = 0\ ,
 \\
 \label{bgr7}
  && \hat d_t \bbox{g}_2+{\hat{\bbox{d}}}\times\bbox{g}_3
     +2{\hat{\bbox{\pi}}}f_3 - 2 m \bbox{g}_1 = 0\ ,
 \\
 \label{bgr8}
  && \hat d_t \bbox{g}_3-{\hat{\bbox{d}}}\times\bbox{g}_2
     -2{\hat{\bbox{\pi}}}f_2 = 0\ .
 \end{eqnarray}
 \end{mathletters}
These $16$ equations are identical with the BGR equations derived by 
Bialynicki-Birula, Gornicki and Rafelski \cite{BGR} by Wigner
transforming the equations of motion for the equal-time density 
matrix. They determine the dynamics of the zeroth-order energy moments.
The three-dimensional dynamical operators occurring in these equations
are identical with the ones arising in scalar QED and are given in 
Appendix~\ref{appc}.  

The first-order moments $f_i^1(x,\bbox{p})$ and $\bbox{g}_i^1(x,\bbox{p})$ 
satisfy $16$ transport equations derived \cite{ZH1} from the first energy 
moment of the covariant transport equations (\ref{spinor1}), 
 \begin{mathletters}
 \label{spintran}
 \begin{eqnarray}
 \label{spintran1}
  && {1\over\sqrt 3}\hat d_t f_0^1 
   + {1\over \sqrt 3}{\hat{\bbox{d}}}{\cdot}\bbox{g}_1^1
   + \hat Df_0+\bbox{I}{\cdot}\bbox{g}_1 = 0\ ,
 \\
 \label{spintran2}
  && {1\over\sqrt 3}\hat d_t f_1^1 + 
     {1\over \sqrt 3}{\hat{\bbox{d}}}{\cdot}\bbox{g}_0^1
     + {2\over \sqrt 3}m f_2^1+\hat Df_1+\bbox{I}{\cdot}\bbox{g}_0 = 0\ ,
 \\
 \label{spintran3}
  && {1\over\sqrt 3}\hat d_t f_2^1  
     +{2\over \sqrt 3}{\hat{\bbox{\pi}}}{\cdot}\bbox{g}_3^1
     -{2\over \sqrt 3}m f_1^1
     + \hat Df_2+ 2{\hat{\bbox{G}}}{\cdot}\bbox{g}_3 = 0\ ,
 \\
 \label{spintran4}
  && {1\over\sqrt 3}\hat d_t f_3^1 
     -{2\over \sqrt 3}{\hat{\bbox{\pi}}}{\cdot}\bbox{g}_2^1
     + \hat Df_3 - 2{\hat{\bbox{G}}}{\cdot}\bbox{g}_2 = 0\ ,
 \\
 \label{spintran5}
  && {1\over \sqrt 3}\hat d_t\bbox{g}_0^1
     +{1\over \sqrt 3}{\hat{\bbox{d}}}f_1^1
     +{2\over \sqrt 3}{\hat{\bbox{\pi}}}\times\bbox{g}_1^1
     +\hat D\bbox{g}_0+{\hat{\bbox{I}}}f_1
     -2{\hat{\bbox{G}}} \times \bbox{g}_1 = 0\ ,
 \\
 \label{spintran6}
  && {1\over \sqrt 3}\hat d_t\bbox{g}_1^1
     +{1\over \sqrt 3}{\hat{\bbox{d}}}f_0^1
     -{2\over \sqrt 3}{\hat{\bbox{\pi}}}\times\bbox{g}_0^1
     +{2\over \sqrt 3}m\bbox{g}_2^1     
     +\hat D\bbox{g}_1+{\hat{\bbox{I}}}f_0
     -2{\hat{\bbox{G}}}\times \bbox{g}_0 = 0\ ,
 \\
 \label{spintran7}
  && {1\over \sqrt 3}\hat d_t\bbox{g}_2^1
     +{1\over \sqrt 3}{\hat{\bbox{d}}}\times \bbox{g}_3^1
     +{2\over \sqrt 3}{\hat{\bbox{\pi}}}f_3^1
     -{2\over \sqrt 3}m \bbox{g}_1^1
     +\hat D\bbox{g}_2+{\hat{\bbox{I}}}\times \bbox{g}_3
     +2{\hat{\bbox{G}}}f_3 = 0\ , 
 \\
 \label{spintran8}
  && {1\over \sqrt 3}\hat d_t\bbox{g}_3^1
     -{1\over \sqrt 3}{\hat{\bbox{d}}}\times\bbox{g}_2^1
     -{2\over \sqrt 3}{\hat{\bbox{\pi}}}f_2^1
     +\hat D\bbox{g}_3-{\hat{\bbox{I}}}\times \bbox{g}_2
     -2{\hat{\bbox{G}}}f_2 = 0\ ,
 \end{eqnarray}
 \end{mathletters}
and $16$ constraint equations derived \cite{ZH1} from the zeroth energy 
moment of the covariant constraint equation (\ref{spinor2}):
 \begin{mathletters}
 \label{spincons}
 \begin{eqnarray}
 \label{spincons1}
  && {1\over \sqrt 3}f_0^1 
     = {\hat{\bbox{\pi}}}{\cdot}\bbox{g}_1 - \hat\pi_0 f_0 + m f_3 \ ,
 \\
 \label{spincons2}
  && {1\over \sqrt 3}f_1^1
     = {\hat{\bbox{\pi}}}{\cdot}\bbox{g}_0 - \hat\pi_0 f_1 \ ,
 \\
 \label{spincons3}
  && {1\over \sqrt 3}f_2^1
     = -{\hbar\over 2}{\hat{\bbox{d}}}{\cdot}\bbox{g}_3
       -\hat\pi_0 f_2 \ ,
 \\
 \label{spincons4}
  && {1\over \sqrt 3}f_3^1
     = {\hbar\over 2}{\hat{\bbox{d}}}{\cdot}\bbox{g}_2
      - \hat\pi_0 f_3 + m f_0 \ ,
 \\
 \label{spincons5}
  && {1\over \sqrt 3}\bbox{g}_0^1
    = {\hbar\over 2}{\hat{\bbox{d}}}\times\bbox{g}_1
     + {\hat{\bbox{\pi}}} f_1 - \hat\pi_0 \bbox{g}_0 + m\bbox{g}_3 \ ,
 \\
 \label{spincons6}
  && {1\over \sqrt 3}\bbox{g}_1^1
    = {\hbar\over 2}{\hat{\bbox{d}}}\times\bbox{g}_0
     + {\hat{\bbox{\pi}}} f_0 - \hat\pi_0 \bbox{g}_1 \ ,
 \\
 \label{spincons7}
  && {1\over \sqrt 3}\bbox{g}_2^1
    = -{\hbar\over 2}{\hat{\bbox{d}}}f_3 
      +{\hat{\bbox{\pi}}} \times \bbox{g}_3
      -\hat\pi_0\bbox{g}_2 \ ,
 \\
 \label{spincons8}
  && {1\over \sqrt 3}\bbox{g}_3^1
    = {\hbar\over 2}{\hat{\bbox{d}}}f_2 
     - {\hat{\bbox{\pi}}} \times \bbox{g}_2
     - \hat\pi_0 \bbox{g}_3 + m \bbox{g}_0 \ .
 \end{eqnarray}
 \end{mathletters}
A discussion similar to that in scalar QED reveals \cite{ABZH} that 
the transport equations (\ref{spintran}) are not independent of the 
BGR equations (\ref{bgr}) and the contraint equations (\ref{spincons}).
For instance, using Eqs.~(\ref{spincons}), the transport equation for 
$f_0^1$ can be expressed as an operator combination of the transport 
equations for $f_0$ and $\bbox{g}_1$: 
 \begin{equation}
    \hat\pi_0\left(\hat d_t f_0 + {\hat{\bbox{d}}}{\cdot}\bbox{g}_1\right)
   -{\hat{\bbox{\pi}}}{\cdot}\left(\hat d_t\bbox{g}_1
   +{\hat{\bbox{d}}}f_0-2{\hat{\bbox{\pi}}}\times \bbox{g}_0
   +2m\bbox{g}_2\right) = 0\ .
 \end{equation}
Therefore, the first-order moments are fully determined in terms of
the solutions of the BGR equations (\ref{bgr}) for the zeroth-order
moments by solving the constraint equations (\ref{spincons}). 

Again, we have not been able to find a simple proof that the same is 
generally true for all higher order energy moments, and we stopped here.
We do, however, believe that such a proof must exist \cite{ochs}, and
that therefore all higher order energy moments can be directly 
computed from the solutions of the BGR transport equations by solving 
the constraint hierarchy.

It was shown in Ref.~\cite{ZH1} that in the classical limit the simple 
algebraic relation (\ref{class2}) changes the structure of the 
constraints (\ref{spincons}) for the first-order moments and turns 
them into additional constraints for the zeroth-order moments (i.e. 
the equal-time Wigner functions). These extra constraints reduce the 
number of independent zeroth-order moments from $16$ in the quantum 
case to $4$ in the classical limit. They are thus extremely important. 
As a result the BGR equations reduce to two decoupled Vlasov-type 
transport equations for the charge and spin distribution functions. In 
the general quantum case there are no such extra constraints on the 
equal-time Wigner functions \cite{ABZH}. One must solve all 16 coupled
transport equations (\ref{bgr}), but these solutions then fully 
determine also all higher order moments. These higher order moments
have important physical meaning: the first-order moment of $F_0(x,p)$, 
for instance, describes the energy distribution in phase-space. With 
the help of the contraint (\ref{spincons1}) it is given by 
 \begin{eqnarray}
   \epsilon(\bbox{x},\bbox{p},t) 
   &=& {\rm Tr} \int {dE\over 2\pi} \, E \, \gamma_0 {\cal W}(x,p)
     = \sqrt{2\over 3}\, f_0^1(\bbox{x},\bbox{p},t)
 \nonumber\\
   &=& \sqrt 2\Bigl( m f_3(\bbox{x},\bbox{p},t) 
     - \hat\pi_0 f_0(\bbox{x},\bbox{p},t) 
     + {\hat{\bbox{\pi}}}{\cdot}\bbox{g}_1(\bbox{x},\bbox{p},t)\Bigr)\, .
 \end{eqnarray}

\section{Conclusions}
\label{concl}

We have presented a universal method for the construction of 
equal-time quantum transport theories from the covariant quantum field
equations of motion. It is based on energy averaging the covariant
kinetic equations for the covariant Wigner operator (which is the 
Wigner transform of the covariant, ``two-time" density matrix)
and its energy moments. This procedure yields a hierarchy of coupled
transport and constraint equations for the energy moments of the 
covariant Wigner function, the so-called equal-time Wigner functions.
We showed how, in the mean-field approximation, this hierarchy can be 
truncated at a rather low level, requiring the solution of only a 
small number of equal-time transport equations, and how the higher 
order energy moments (higher order equal-time Wigner functions) can be 
constructed from these solutions via constraints.  

The major advantage of the equal-time formulation of (quantum) 
transport theory is that the resulting transport equations can be 
solved as initial value problems, with boundary values for the 
equal-time Wigner functions at $t=-\infty$ which can be calculated 
from the fields at $t=-\infty$. This is not the case for the covariant 
transport equations and the covariant Wigner function. The present 
paper thus provides an essential step in the direction of practical 
computations of the dynamics of relativistic quantum field systems out
of thermal equilibrium in the language of transport theory, i.e. in a 
phase-space oriented approach. The method presented here improves upon 
previous approaches by being much more systematic: we did not just focus
on the lowest energy moments (which contain only a small fraction 
of the information contained in the covariant Wigner function), but 
we discussed and showed how to solve the complete hierarchy of moment 
equations. We had already before demonstrated for spinor QED that the 
non-covariant three-dimensional approach (which starts directly from the 
equal-time density matrix) yields an incomplete set of equal-time 
transport equations. In this paper we also discussed the case of 
scalar QED and again discovered serious problems with the direct 
three-dimensional approach based on the Feshbach-Villars formulation.
We conclude that the only safe way of deriving a correct and complete 
set of equal-time quantum transport equations is by starting from the 
covariant formulation and taking energy moments of the covariant 
kinetic equations. The method can be generalized in a straightforward 
way to other types of interactions \cite{SR,ZH2,ABZH}, including 
non-Abelian gauge interactions \cite{ochs}.

The structure of the hierarchy of equal-time quantum kinetic equations 
depends on the structure of the covariant field equations from which 
one starts. For scalar or vector theories with second order time 
derivatives one has to solve a coupled set of three equations for the 
three lowest energy moments, two resulting from the equal-time 
transport hierarchy and one stemming from the constraint hierarchy.  
For spinor theories with only first order time derivatives in the 
field equations one ends up with only one equal-time transport 
equation for the lowest energy moment of each spinor component of the 
Wigner function. All higher order energy moments can be determined 
from the solutions of these equations by solving constraints.  

Important further simplifications occur in the classical limit 
$\hbar\to 0$: then all higher order energy moments can be expressed
algebraically in terms of the zeroth energy moment, and the number of 
equations is drastically reduced. For scalar theories one obtains just 
one Vlasov-type equation for the on-shell charge distribution 
function. For spinor theories one obtains two decoupled Vlasov-type 
equations (one scalar and one vector equation) for the on-shell charge 
and spin density distributions in phase-space. Again the only 
systematic way of deriving the constraints leading to these 
simplifications is by energy averaging the (classical limit of the) 
covariant transport equations.  

All results in this paper were derived in the mean field 
approximation, i.e. in the collisionless limit. It is generally known 
that including collision terms in the covariant transport equations 
leads to the appearance of non-localities in time (``memory effects")
in the equal-time transport equations \cite{Mal}. It is not inconceivable 
that these memory effects lead to serious complications for the truncation 
of the equal-time transport hierarchy. This is certainly an 
interesting and difficult problem for future studies.

\acknowledgments

P.Z. thanks GSI for a fellowship. The work of U.H. was supported 
by BMBF, DFG, and GSI. He would like to express his thanks to CERN, 
where this work was completed, for warm hospitality and a stimulating 
environment.  

\appendix
\section{The coefficients $\lowercase{c^{mn}_{ij}}$}
\label{appa}

>From the definition of the coefficients $c_{ij}^{mn}$ in 
(\ref{hierarchy2b}) and the orthogonality condition  (\ref{ortho}) we 
see that 
 \begin{equation}
 \label{a1}
   c_{ij}^{00} = \delta_{ij} \, .
 \end{equation}
Furthermore, it is easy to see from (\ref{hierarchy2b}) that 
 \begin{equation}
 \label{a2}
   c_{ij}^{mn} = 0 \qquad \text{for} \qquad n>i+m \qquad \text{or} \qquad
                 \ j>i+m-n\ . 
 \end{equation}
Therefore, if the finite sum over $m$ has the upper limit $M$, the sum 
over $j$ in Eq.~(\ref{hierarchy2}) extends only over the range 
$j \leq i+M$.

>From the recursion relation for the Legendre polynomials 
 \begin{equation}
 \label{recur1}
    \omega P_i(\omega) = {i+1 \over 2i+1} P_{i+1}(\omega) +
    {i\over 2i+1} P_{i-1}(\omega)
 \end{equation}
one easily derives the recursion relation
 \begin{equation}
 \label{R1}
    c^{m+1,n}_{i,j} = {i+1 \over \sqrt{(2i+1)(2i+3)}}\, c^{m,n}_{i+1,j} +
    {i\over \sqrt{(2i+1)(2i-1)}}\, c^{m,n}_{i-1,j}
 \end{equation}
for the coefficients $c^{mn}_{ij}$. This allows to raise the first 
upper index $m$, starting from (\ref{a1}). In particular we have
 \begin{mathletters}
 \label{cij}
 \begin{eqnarray}
 \label{cij1}
    c^{10}_{i-1,i} &=& {i \over \sqrt{(2i-1)(2i+1)}}\, ,
 \\
 \label{cij2}
    c^{20}_{i-1,i+1} &=& {i (i+1)\over (2i+1) \sqrt{(2i-1)(2i+3)}}\, ,
 \end{eqnarray}
 \end{mathletters}
which we will need in Appendix~\ref{appb}. Similarly the second upper 
index can be raised by using the relations
 \begin{equation}
 \label{recur3}
   \partial_\omega \Bigl( P_{j+1}(\omega) - P_{j-1}(\omega) \Bigr)
   = (2j+1) P_j(\omega)\, ,\qquad
   P_j(\pm 1) = (\pm 1)^j\, \quad(j\geq 0)\, ,  
 \end{equation}
which, for $j\geq 2$, lead to
 \begin{equation}
 \label{R2}
    c^{m,n+1}_{i,j} = \sqrt{2j+1\over 2j-3}\, c^{m,n+1}_{i,j-2} 
                      + \sqrt{(2j+1)(2j-1)}\, c^{m,n}_{i,j-1} \, .
 \end{equation}
This expression is useless for $j=0$ and $j=1$; these cases can be treated 
by another recursion relation which can be obtained by using (\ref{recur3})
on the Legendre polynomial with the index $i$:
 \begin{equation}
 \label{R3}
    c^{m,n+1}_{i,j} = \sqrt{2i+1\over 2i-3}\, c^{m,n+1}_{i-2,j} 
                      - \sqrt{(2i+1)(2i-1)}\, c^{m,n}_{i-1,j} 
        -m \left( c^{m-1,n}_{i,j} - 
                  \sqrt{2i+1\over 2i-3}\, c^{m-1,n}_{i-2,j}
           \right) \, .
 \end{equation}
For $i<2$ this must be used together with
 \begin{equation}
 \label{recur4}
    c^{m,n}_{-i,j} = \sqrt{-1}\, c^{m,n}_{i-1,j}
 \end{equation}
which follows from $P_{-i}(x) = P_{i-1}(x)$.

The above recursion relations can be initialized with the following
nonvanishing coefficients for $i,j,m,n\leq 1$:
 \begin{equation}
 \label{a4}
   c^{00}_{00} = c^{00}_{11} =  - c^{11}_{00} = 1\, ,  \ \
   c^{10}_{01} = c^{10}_{10} = {1\over \sqrt{3}}\, , \ \
   c^{01}_{10} = - \sqrt{3}\, ,\ \ c^{11}_{11} = -2\, . 
 \end{equation}

\section{The operators $\lowercase{\hat g_{ij}}$ and 
         $\lowercase{\hat f_{ij}}$ for scalar theories}
\label{appb}

The three-dimensional dynamical operators $\hat g_{ij}(x,\bbox{p})$ and 
$\hat f_{ij}(x,\bbox{p})$ needed in Eqs.~(\ref{mini}) and (\ref{next}) 
are obtained from the definition (\ref{hierarchy2a}) with the 
coefficients $c_{ij}^{mn}$ from Appendix~\ref{appa}. Please note that
for $\hat g_{ij}$ the upper limit in (\ref{hierarchy2a}) for the sum 
over $m$ is $M=1$ while for $\hat f_{ij}$ it is $M+1=2$. One finds:
 \begin{mathletters}
 \label{gfij}
 \begin{eqnarray}
 \label{g00}
   \hat g_{00} &=& \hat G_{00}-\hat G_{11}\ ,
 \\
 \label{g01}
   \hat g_{01} &=& {1\over \sqrt 3}\hat G_{10}\ ,
 \\
 \label{g10}
   \hat g_{10} &=& \sqrt 3 \left(-\hat G_{01}+{1\over 3}\hat G_{10}
                                 +2\hat G_{12} \right)\ ,
 \\
 \label{g11}
   \hat g_{11} &=& \hat G_{00}-2\hat G_{11}\ ,
 \\ 
 \label{g12}
   \hat g_{12} &=& {2\over \sqrt{15}}\hat G_{10}\ ,
 \\
 \label{g20}
   \hat g_{20} &=& \sqrt 5 \left( 3\hat G_{02}-\hat G_{11}-9\hat G_{13}
                           \right) \ ,
 \\
 \label{g21}
   \hat g_{21} &=& \sqrt{15} \left(-\hat G_{01}+{2\over 15} \hat G_{10}
                   +3\hat G_{12}\right)\ ,
 \\
 \label{g22}
   \hat g_{22} &=& \hat G_{00}-3\hat G_{11}\ ,
 \\
 \label{g23}
   \hat g_{23} &=& {3\over \sqrt{35}}\hat G_{10}\ ,
 \\
 \label{f00}
   \hat f_{00} &=& \hat F_{00}-\hat F_{11} + {1\over 3} \hat F_{20}
                              + 2 \hat F_{22}\ ,
 \\
 \label{f01}
   \hat f_{01} &=& {1\over \sqrt 3} \left(\hat F_{10} - 2\hat F_{21}\right)\ ,
 \\
 \label{f02}
   \hat f_{02} &=& {2\over 3\sqrt 5}\hat F_{20}\ ,
 \\
 \label{f10}
   \hat f_{10} &=& \sqrt 3 \left(-\hat F_{01}+{1\over 3}\hat F_{10}
                   +2\hat F_{12}-\hat F_{21}-6\hat F_{23} \right)\ ,
 \\
 \label{f11}
   \hat f_{11} &=& \hat F_{00}-2\hat F_{11} + {3\over 5} \hat F_{20}
               + 6 \hat F_{22}\ ,
 \\
 \label{f12}
   \hat f_{12} &=& {2\over \sqrt{15}}\left(\hat F_{10}-3\hat F_{21}\right)\ ,
 \\
 \label{f13}
   \hat f_{13} &=& {2\sqrt 3\over 5\sqrt{7}} \hat F_{20}\ .
 \end{eqnarray}
 \end{mathletters}
The dynamical operators $\hat G_{mn}$ and $\hat F_{mn}$ are given
in equations (\ref{gfmn}) for scalar mean field interactions and 
in Appendix~\ref{appc} for scalar QED. Please note that in both cases
$\hat F_{2k} = 0$ for $k\ne 0$ and thus some of the expressions for $\hat
f_{ij}$ above simplify. 

We also note that in
 \begin{equation}
 \label{fii}
   \hat f_{i-1,i+1} = \sum_{m=0}^2 \sum_{n=0}^{i-1+m} c_{i-1,i+1}^{mn} 
   \hat F_{mn}
 \end{equation}
the only term which doesn't violate the second inequality in (\ref{a2}) 
is the one with $m=2,n=0$:
 \begin{equation}
 \label{fii1}
   \hat f_{i-1,i+1} = c_{i-1,i+1}^{20} \hat F_{20}
                    = {i(i+1) \over (2i+1) \sqrt{(2i-1)(2i+3)}} \hat F_{20}\, .
 \end{equation}
In the last equality we used (\ref{cij2}). From Eqs.~(\ref{fmn}) or 
(\ref{C4g}) we thus see that $\hat f_{i-1,i+1}$ is just a constant.
Following the same reasoning one also obtains
 \begin{equation}
 \label{gii}
   \hat g_{i-1,i} = \sum_{m=0}^1 \sum_{n=0}^{i+m} c_{i-1,i}^{mn} \hat G_{mn}
                  =  c_{i-1,i}^{10} \hat G_{10}
                  = {i \over \sqrt{(2i-1)(2i+1)}} \hat G_{10}
   \, ,
 \end{equation}
where according to Eqs.~(\ref{gmn}) or (\ref{C3d}) $\hat G_{10}$ is 
proportional to the time derivative $\partial_t$ resp. $\hat d_t$.

\section{ The operators $\hat G_{\lowercase{mn}}$ and 
          $\hat F_{\lowercase{mn}}$ for scalar QED }
\label{appc}
 
The elementary operators $\hat \Pi_\mu$ and $\hat D_\mu$ are extensions 
of the covariant momentum $p_\mu$ and the covariant derivative 
$\partial_\mu$, respectively. Their double expansions in $\omega$ and 
$\partial / \partial \omega$ are
 \begin{mathletters}
 \label{C1}
 \begin{eqnarray}
 \label{C1a}
   \hat\Pi_0 &=& \Lambda\omega+\hat\pi_0
                  +{\hat A\over\Lambda}{\partial\over\partial \omega}
                  -{\hat B\over \Lambda^2}{\partial^2\over \partial \omega^2}
                  -{\hat C\over \Lambda^3}{\partial^3\over \partial \omega^3}
                  +\cdots\ ,
 \\
 \label{C1b}
   \hat D_0   &=& \hat d_t-{\hat D\over \Lambda}{\partial\over \partial \omega}
                  -{\hat E\over \Lambda^2}{\partial^2\over \partial \omega^2}
                  +{\hat F\over \Lambda^3}{\partial^3\over \partial \omega^3}
                  +\cdots\ ,
 \\
 \label{C1c}
   {\hat{\bbox{\Pi}}} &=& {\hat{\bbox{\pi}}}
          -{{\hat{\bbox{G}}}\over\Lambda}{\partial\over \partial \omega}
          +{{\hat{\bbox{H}}}\over\Lambda^2}{\partial^2\over\partial\omega^2}
                  +\cdots ,
 \\
 \label{C1d}
   {\hat{\bbox{D}}} &=& -{\hat{\bbox{d}}}
        +{{\hat{\bbox{I}}}\over\Lambda}{\partial\over\partial \omega}
        +{{\hat{\bbox{J}}}\over\Lambda^2}{\partial^2\over \partial \omega^2}
        +\cdots\ ,
 \end{eqnarray}
 \end{mathletters}
with the equal-time dynamical operators,
 \begin{mathletters}
 \label{C2}
 \begin{eqnarray}
 \label{C2a}
   \hat\pi_0 (x,\bbox{p}) &=& i e \hbar \int^{1\over 2}_{-{1\over 2}}ds\, s\, 
                \bbox{E}(\bbox{x} + i s \hbar \bbox{\nabla}_{\!p},t)
                             {\cdot}\bbox{\nabla}_{\!p} \ ,
 \\
 \label{C2b}
   {\hat{\bbox{\pi}}}(x,\bbox{p}) &=& \bbox{p}-i e \hbar \int^{1\over 2}_
                                {-{1\over 2}}ds\, s\,\bbox{B}(\bbox{x} 
                                + is\hbar\bbox{\nabla}_{\!p}, t)
                                \times \bbox{\nabla}_{\!p} \ ,
 \\
 \label{C2c}
   \hat d_t(x,\bbox{p}) &=& \hbar\partial_t + e \hbar\int^{1\over 2}_
          {-{1\over 2}} ds\,\bbox{E}(\bbox{x}+is\hbar
          \bbox{\nabla}_{\!p}, t){\cdot}\bbox{\nabla}_{\!p} \ ,
 \\
 \label{C2d}
   {\hat{\bbox{d}}}(x,\bbox{p}) 
   &=& \hbar\bbox{\nabla}_{\!x} + e \hbar\int^{1\over 2}_
      {-{1\over 2}}ds\,\bbox{B}(\bbox{x}+is\hbar \bbox{\nabla}_{\!p}, t) 
      \times \bbox{\nabla}_{\!p}\ ,
 \\
 \label{C2e}
   \hat A(x,\bbox{p}) &=& e\hbar^2 \int^{1\over 2}_{-{1\over 2}}ds\, s^2 
                         {\partial \over \partial t}\bbox{E}(\bbox{x}+is\hbar
                         \bbox{\nabla}_{\!p}, t){\cdot}\bbox{\nabla}_{\!p} \ ,
 \\
 \label{C2f}
   \hat B(x,\bbox{p}) &=& {ie\hbar^3\over 2}\int^{1\over 2}_{-{1\over 2}}ds\, 
         s^3 {\partial^2 \over \partial t^2}\bbox{E}(\bbox{x}+
         is\hbar\bbox{\nabla}_{\!p}, t){\cdot}\bbox{\nabla}_{\!p} \ ,
 \\
 \label{C2g}
   \hat C(x,\bbox{p}) &=& {e\hbar^4\over 6}\int^{1\over 2}_{-{1\over 2}}ds\, 
         s^4 {\partial^3 \over \partial t^3}\bbox{E}(\bbox{x}+
         is\hbar\bbox{\nabla}_{\!p}, t){\cdot}\bbox{\nabla}_{\!p} \ ,
 \\
 \label{C2h}
   \hat D(x,\bbox{p}) &=& ie\hbar^2 \int^{1\over 2}_{-{1\over 2}}ds\, s 
         {\partial \over \partial t}\bbox{E}(\bbox{x}
         +is\hbar\bbox{\nabla}_{\!p}, t){\cdot}\bbox{\nabla}_{\!p} \ ,
 \\
 \label{C2i}
   \hat E(x,\bbox{p}) &=& {e\hbar^3\over 2} \int^{1\over 2}_{-{1\over 2}}ds\, 
         s^2 {\partial^2 \over \partial t^2}\bbox{E}(\bbox{x}
         +is\hbar\bbox{\nabla}_{\!p}, t){\cdot}\bbox{\nabla}_{\!p} \ ,
 \\
 \label{C2j}
   \hat F(x,\bbox{p}) 
   &=& {ie\hbar^4\over 6} \int^{1\over 2}_{-{1\over 2}}ds\, 
       s^3 {\partial^3 \over \partial t^3}\bbox{E}(\bbox{x}
       +is\hbar\bbox{\nabla}_{\!p}, t){\cdot}\bbox{\nabla}_{\!p} \ ,
 \\
 \label{C2k}
  {\hat{\bbox{G}}}(x,\bbox{p}) 
  &=& e\hbar^2 \int^{1\over 2}_{-{1\over 2}} ds\, s^2
      {\partial\over\partial t}\bbox{B}(\bbox{x}
      +is\hbar\bbox{\nabla}_{\!p}, t)\times \bbox{\nabla}_{\!p}
      +ie\hbar\int^{1\over 2}_{-{1\over 2}}ds\, s 
      \bbox{E}(\bbox{x}+is\hbar\bbox{\nabla}_{\!p},t)\ ,
 \\
 \label{C2l}
   {\hat{\bbox{H}}}(x,\bbox{p}) 
   &=& {ie\hbar^3\over 2} \int^{1\over 2}_
   {-{1\over 2}} ds\, s^3{\partial^2\over\partial t^2}\bbox{B}(\bbox{x}
    +is\hbar\bbox{\nabla}_{\!p}, t)\times \bbox{\nabla}_{\!p}
 \nonumber\\
   &&-e\hbar^2\int^{1\over 2}_{-{1\over 2}}ds\, s^2 
    {\partial \over\partial t} \bbox{E}(\bbox{x}
    +is\hbar\bbox{\nabla}_{\!p},t) ,
 \\
 \label{C2m}
   {\hat{\bbox{I}}}(x,\bbox{p}) 
   &=& ie\hbar^2\int^{1\over 2}_{-{1\over 2}} ds\, s
       {\partial\over\partial t}\bbox{B}(\bbox{x}
       +is\hbar\bbox{\nabla}_{\!p}, t)\times \bbox{\nabla}_{\!p}
       -e\hbar\int^{1\over 2}_{-{1\over 2}}ds 
       \bbox{E}(\bbox{x}+is\hbar\bbox{\nabla}_{\!p},t)\ ,
 \\
 \label{C2n}
   {\hat{\bbox{J}}}(x,\bbox{p}) 
   &=& {e\hbar^3\over 2}\int^{1\over 2}_{-{1\over 2}} 
       ds\, s^2{\partial^2\over\partial t^2}\bbox{B}(\bbox{x}
           +is\hbar\bbox{\nabla}_{\!p}, t)\times \bbox{\nabla}_{\!p}
 \nonumber\\
   && +ie\hbar^2\int^{1\over 2}_{-{1\over 2}}ds\, s 
       {\partial \over \partial t}\bbox{E}(\bbox{x}
       +is\hbar\bbox{\nabla}_{\!p},t) .
  \end{eqnarray}
 \end{mathletters}
By substituting these expansions into the expressions of
$\hat G$ and $\hat F$ in (\ref{gf}) and using the definition of
$\hat G_{mn}$ (\ref{gradient}), we list some low-order $\hat G_{mn}$ and 
$\hat F_{mn}$ which 
will be used in the derivation of the kinetic equations for low-order
energy moments,
 \begin{mathletters}
 \label{C3}
 \begin{eqnarray}
 \label{C3a} 
   \hat G_{00} &=& \hat\pi_0\hat d_t 
   +{\hat{\bbox{\pi}}}{\cdot}{\hat{\bbox{d}}}\ ,
 \\
 \label{C3b} 
   \hat G_{01} &=& -{1\over \Lambda}\left(\hat\pi_0\hat D 
   +{\hat{\bbox{\pi}}}{\cdot}{\hat{\bbox{I}}}
   -\hat A \hat d_t+{\hat{\bbox{G}}}{\cdot}{\hat{\bbox{d}}}\right)\ ,
 \\
 \label{C3c} 
   \hat G_{02} &=& {1\over \Lambda^2}\left(
                   {\hat{\bbox{H}}}{\cdot}{\hat{\bbox{d}}}
                   +{\hat{\bbox{G}}}{\cdot}{\hat{\bbox{I}}}
                   -{\hat{\bbox{\pi}}}{\cdot}{\hat{\bbox{J}}}
                   -\hat B\hat d_t-\hat A\hat D
                   -\hat\pi_0 \hat E \right)\ ,
 \\
 \label{C3d} 
   \hat G_{10} &=& \Lambda\hat d_t\ ,
 \\
 \label{C3e} 
   \hat G_{11} &=& -\hat D\ ,
 \\
 \label{C3f} 
   \hat G_{12} &=& -{\hat E\over \Lambda}\ ,
 \\
 \label{C3g} 
   \hat G_{13} &=& {\hat F\over \Lambda^2}\ ,
 \end{eqnarray}
 \end{mathletters}
and
 \begin{mathletters}
 \label{C4}
 \begin{eqnarray}
 \label{C4a} 
   \hat F_{00} &=& {1\over 4}(\hat d_t^2-{\hat{\bbox{d}}}^2)
   -(\hat A+\hat\pi_0^2 -{\hat{\bbox{\pi}}}^2 ) +m^2\, ,
 \\
 \label{C4b} 
   \hat F_{01} &=& {1\over \Lambda}\left(
                   {1\over 4}( \{{\hat{\bbox{d}}},{\hat{\bbox{I}}}\}
                   -\{\hat d_t,\hat D\})
                   -2\hat\pi_0\hat A
                   -\{{\hat{\bbox{\pi}}},{\hat{\bbox{G}}}\}
                   +2\hat B\right)\, ,
 \\
 \label{C4c} 
   \hat F_{02} &=& {1\over \Lambda^2}\left(
                   {1\over 4}\Bigl(\hat D^2 - {\hat{\bbox{I}}}^2
                   +\{ {\hat{\bbox{d}}},{\hat{\bbox{J}}} \}
                   -\{ \hat d_t,\hat E \} \Bigr) 
                   -\Bigl(\hat A^2-{\hat{\bbox{G}}}^2
                   -\{{\hat{\bbox{\pi}}},{\hat{\bbox{H}}}\}
                   -2\hat\pi_0\hat B \Bigr)
                   +3\hat C\right) ,
 \\
 \label{C4d} 
   \hat F_{10} &=& -2\Lambda\hat\pi_0\, ,
 \\
 \label{C4e} 
   \hat F_{11} &=& -2\hat A\, ,
 \\
 \label{C4f} 
   \hat F_{12} &=& 2{\hat B\over\Lambda}\, ,
 \\
  \label{C4g} 
   \hat F_{20} &=& -\Lambda^2\, .
 \end{eqnarray}
 \end{mathletters}

%
%


\begin{thebibliography}{99}
  \bibitem[*]{home}  On leave from Hua-Zhong Normal University, Wuhan, China.
  
  \bibitem{EH}   {H.-Th. Elze and U. Heinz, 
                  Phys. Rep. {\bf 183}, 81 (1989).}
  \bibitem{CZ}   {P. Carruthers and F. Zachariasen, 
                  Rev. Mod. Phys. {\bf 55}, 245 (1983).} 
  \bibitem{He}   {U. Heinz, 
                  Phys. Rev. Lett. {\bf 51}, 351 (1983).}
  \bibitem{EGV1} {H.-Th. Elze, M. Gyulassy and D. Vasak, 
                  Nucl. Phys. {\bf B276}, 706 (1986).}
  \bibitem{EGV2} {H.-Th. Elze, M. Gyulassy and D. Vasak, 
                  Phys. Lett.{\bf B177}, 402 (1986).}
  \bibitem{VGE}  {D. Vasak, M. Gyulassy and H.-Th. Elze, 
                  Ann. Phys. (N.Y.) {\bf 173}, 462 (1987).}  
  \bibitem{BGR}  {I. Bialynicki-Birula, P. Gornicki and J. Rafelski,
                  Phys. Rev. {\bf D44}, 1825 (1991).} 
  \bibitem{BGG}  {C. Best, P. Gornicki and W. Greiner, 
                  Ann. Phys. (N.Y.) {\bf 225}, 169(1993).}
  \bibitem{Hen}  {P. Henning, 
                  Phys. Rep. {\bf 253}, 263 (1995).}
  \bibitem{Sch}  {J. Schwinger, 
                  Phys. Rev. {\bf 82}, 664 (1951);\\ 
                  Y. Kluger, J. M. Eisenberg, B. Svetitsky, F. Cooper 
                  and E. Mottola, 
                  Phys. Rev. Lett. {\bf 67}, 2427 (1991);
                  and Phys. Rev. {\bf D45}, 4659 (1992).} 
  \bibitem{BE}   {C. Best and J. M. Eisenberg, 
                  Phys. Rev. {\bf D47}, 4639 (1993).} 
  \bibitem{ZH1}  {P. Zhuang and U. Heinz, 
                  Ann. Phys. (N.Y.) {\bf 245}, 311 (1996);\\
                  and {\it ibid.}, Erratum {\bf 249}, 362 (1996).}
  \bibitem{GLW}  {S. R. deGroot, W. A. van Leeuwen, and Ch. G. van Weert,
                  ``Relativistic Kinetic Theory'', NorthHolland, Amsterdam,
                  1980.}
  \bibitem{FV}   {H. Feshbach and F. Villars, 
                  Rev. Mod. Phys. {\bf 30}, 24 (1958).}
  \bibitem{Best}  {The last equation (\ref{density4}) was given incorrectly
                   in \cite{BGG} and must be corrected as shown (C. Best, 
                   private communication).}
  \bibitem{SR}    {G. R. Shin and J. Rafelski,
                   Ann. Phys. (N. Y.) {\bf 243}, 65 (1995).} 
  \bibitem{ZH2}   {P. Zhuang and U. Heinz, 
                   Phys. Rev. {\bf D53}, 2096 (1996).}
  \bibitem{ABZH}  {A. Abada, M. C. Birse, P. Zhuang and U. Heinz,
                   Phys. Rev. {\bf D54}, 4175 (1996).}
  \bibitem{ochs}  {We just learned that S. Ochs has apparently found 
                   such a general proof (S. Ochs, PhD thesis, 
                   University of Regensburg, Dec. 1996, unpublished).}
  \bibitem{Mal}   {R. Malfliet, in {\it Correlations and Clustering 
                   Phenomena in Subatomic Physics}, edited by 
                   M.N.~Harakeh, O.~Scholten, and J.H.~Koch, NATO ASI
                   Series (Plenum, New York, 1996), in press.}
\end{thebibliography}
\end{document}